\documentclass[aps,prl,superscriptaddress,showpacs,floatfix,preprint]{revtex4}
\usepackage{amsmath,graphicx,color,amssymb}

\newcommand{\bl}[1]{\begin{equation}\label{#1}}
\newcommand{\be}{\begin{equation}}
\newcommand{\ee}{\end{equation}}
\newcommand{\bea}{\begin{eqnarray}}
\newcommand{\eea}{\end{eqnarray}}

\newcommand{\pd}[2]{\frac{\partial{#1}}{\partial{#2}}}
\newcommand{\td}[2]{\frac{\mathrm{d}{#1}}{\mathrm{d}{#2}}}
\newcommand{\rec}[1]{\frac{1}{#1}}
\newcommand{\z}[1]{\left({#1}\right)}
\newcommand{\sz}[1]{\left[{#1}\right]}
\newcommand{\kz}[1]{\left\{{#1}\right\}}
\renewcommand{\sp}{,\quad\quad}
\renewcommand{\v}[1]{\mathbf{#1}}
\newcommand{\m}[1]{\mathrm{#1}}
\renewcommand{\c}[1]{\mathcal{#1}}

\renewcommand{\r}[1]{(\ref{#1})}
\newcommand{\eq}[1]{eq.~(\ref{#1})}
\newcommand{\eqs}[2]{eqs.~(\ref{#1}) and (\ref{#2})}

\newcommand{\Eq}[1]{Eq.~(\ref{#1})}

\newcommand{\elte}{ELTE, E{\"o}tv{\"o}s Lor{\'a}nd University, H - 1117 Budapest, P{\'a}zm{\'a}ny P. s. 1/A, Hungary}
\newcommand{\kfki}{MTA KFKI RMKI, H-1525 Budapest 114, POBox 49, Hungary}
\newcommand{\stonych}{Department of Chemistry, SUNY Stony Brook, Stony Brook, NY, 11794-3400, USA}

\begin{document}

\title{Detailed description of accelerating, simple solutions of
relativistic perfect fluid hydrodynamics}

\author{M.~I.~Nagy}        \affiliation{\kfki}
\author{T.~Cs{\"o}rg\H{o}}      \affiliation{\kfki}
\author{M.~Csan\'ad}    \affiliation{\elte}\affiliation{\stonych}

\begin{abstract}
In this paper we describe in full details a new family of recently found
exact solutions of relativistic, perfect fluid dynamics. 
With an ansatz, which generalizes the well-known Hwa-Bjorken solution, 
we obtain a wide class of new exact, explicit and simple solutions, 
which have a remarkable advantage as compared to presently known exact and
explicit solutions: they do not lack acceleration. They can
be utilized for the description of the evolution of the
matter created in high energy heavy ion collisions. Because
these solutions are accelerating, they provide a more
realistic picture than the well-known Hwa-Bjorken solution,
and give more insight into the dynamics of the matter. We
exploit this by giving an advanced simple estimation of
the initial energy density of the produced matter in high
energy collisions, which takes acceleration effects (i.e.
the work done by the pressure and the modified change of the volume
elements) into account. We also give an advanced estimation of
the life-time of the reaction. Our new solutions can also 
be used to test numerical hydrodynamical codes reliably.
In the end, we also give an exact, 1+1 dimensional, relativistic
hydrodynamical solution, where the initial pressure and velocity
profile is arbitrary, and we show that this general solution
is stable for perturbations.

\end{abstract}

\maketitle
\date{\today}

\section{Introduction}

Hydrodynamics is in a unique position among the different
branches of theoretical physics: it relies only on the assumption
of local charge and energy-momentum conservation, and on the concept
of local thermal equilibrium. The hydrodynamical equations do not
have an internal scale, in contrast to other theories. Consequently,
among the possible applications of hydrodynamics we find the largest
known physical systems, systems of intermediate size and the smallest
and shortest events of nature as well. For example, it is well known,
that the evolution of the whole Universe (the Hubble flow) can be
treated in a hydrodynamical manner. The applications in the ordinary 
sizes and temporal extents are known since the birth of hydrodynamics.
Considering the smallest temporal and spatial scales: 
as a surprising new result, it has been understood recently, 
that the new form of nuclear matter created
in high energy heavy ion collisions can be considered as an (almost)
perfect fluid --- because its properties (like the single-particle
spectra, elliptic flow data and two-particle correlation functions)
are best explained in terms of models based on perfect fluid
hydrodynamics. From this point of view, we can say that hydrodynamics
is simple and efficient, and it has applications from the very small
to the very large scales.

The hydrodynamical equations, however, are highly nonlinear even in
the nonrelativistic case, this is the reason for the appearance of
chaos, instability and other beautiful flow patterns in realistic
flows. Recent interest in hydrodynamics emerged mostly from the
challenge to explain the properties of the nuclear matter in high
energy collisions, it is clear that in this context one has to deal
with not only hydrodynamics, but relativistic hydrodynamics. Relativity
makes the case even worse: relativistic terms (such as speed-addition
and proper-time factors) make the equations much more complicated
as compared to the nonrelativistic case. From this point of view, hydrodynamics
is a horrendously complicated area of physics, which can be seen also
from the intensity of recent efforts --- both analytical and numerical
--- to find relevant solutions for the relativistic hydrodynamical
equations.

There are only a few exact solutions for these equations. One (and
historically the first) is the famous Landau-Khalatnikov solution
discovered more than 50 years ago~\cite{Landau:1953gs,Khalatnikov,Belenkij:1956cd}.
This is a 1+1 dimensional solution, and has realistic properties:
it describes a 1+1 dimensional expansion, and does not lack acceleration.
Landau was able to calculate approximately an important observable,
the rapidity distribution of the flowing particles from this solution,
and it was found to be an (approximately) Gaussian distribution. Recent
measurements at the RHIC particle accelerator (BNL, USA) suggest that
this prediction is not at big variance with the observations made even at
the highest presently attainable collision energies. However, considering
its formulation, the Landau-Khalatnikov solution is a very unpleasant
one: it is an implicit solution, the independent variables (the space
and and the time coordinate) are given by extremely complicated
integral formulas, involving the fluid rapidity and temperature.

Another renowned relativistic hydrodynamical solution is the
Hwa-Bjorken solution~\cite{Hwa:1974gn,Bjorken:1982qr}, which is
a simple, explicit and exact, but accelerationless solution. This
solution (in its original form, where no conserved baryonic charges
are present) is boost-invariant, and from this it follows that the
rapidity distribution should be constant. This prediction would be
valid for the limiting case of infinitely high center of mass
collision energy. Although such a flat ,,plateaux'' (a narrow domain
of constant rapidity distribution around mid-rapidity) is indeed
seen in recent rapidity measurements at RHIC, the observed distribution
is far from constant, so Bjorken's approximation fails to describe the
data. However, the solution allowed Bjorken to obtain
a simple estimate of the initial energy density reached in high energy
reactions from final state hadronic observables. (This is why mostly
only Bjorken is credited with this solution, though Hwa discovered it
independently almost 10 years before.) The Bjorken-estimation is
widely used, but it is well known, that the acceleration of the matter
influences the energy density estimation. In order to obtain more
realistic estimation, one should use explicit and exact, but
accelerating solutions of relativistic fluid dynamics. 

In this paper we present a  detailed description of such new simple, 
exact and explicit solutions of relativistic fluid dynamics. 
The essential features of these solutions were first presented in two short Letters in 
refs.~\cite{Csorgo:2006ax,Csorgo:2007ea}.
To our best knowledge, these solutions were the first explicit solutions
with relativistic acceleration, following the implicit and accelerating 
Landau-Khalatnikov solution. (There are
other interesting solutions to relativistic hydrodynamics, see
e.g. refs.~\cite{Biro:2000nj,Csorgo:2003rt,Csorgo:2003ry,Sinyukov:2004am},
but these solutions are all accelerationless.)

Apart from the purely theoretical point of view of finding 
new exact solutions for nonlinear differential equations, 
our new solutions are important due to other reasons, too. 
First, they provide an advanced estimate of the initial energy
density and life-time of high energy reactions, we focus on these applications in this
paper. Second, they can be applied to test numerical solutions of
relativistic hydrodynamics: no finite, accelerating
relativistic solutions were available before for 1+3 dimensional tests.

The organization of the paper is as follows. We discuss the equations
of relativistic hydrodynamics, then present the newly found solutions. We then
calculate the rapidity distribution, 
and use it to estimate the initial energy density and the life-time. 
In the four Appendices,  details of calculations are summarized.
In Appendix A we derive the solutions themselves, and in Appendix B we
prove their uniqueness (in some sense discussed there). Appendix C presents the
calculation of the rapidity distribution, and finally, in Appendix D 
we present a general solution and investigate the stability of the solutions
for a specific choice of the equation of state.

\section{Notation and basic equations}

In this section we specify the notations used throughout the paper, and
derive the equations of relativistic hydrodynamics.

\subsection{Notation}

We use three-dimensional as well as four-dimensional notations
depending on which one is more convenient. In four-dimensional
notation the independent variables are the components of the
$x^\mu$ coordinate four-vector: $x^\mu=\z{t,\v{r}}$, with $t$
being the time and $\v{r}$ the coordinate three-vector,
$\v{r}=\z{r_x,r_y,r_z}$. The metric tensor is denoted by
$g_{\mu\nu}$, we use the $g_{\mu\nu}=diag\z{1,-1,-1,-1}$
sign convention. The four-velocity field is denoted by $u^\mu$,
normalized to unity, $u^\mu u_\mu=1$, that is, we treat the
speed of light as $1$. The three-velocity $\v{v}$ is defined
as $u^\mu=\gamma\z{1,\v{v}}$, with $\gamma=\z{1-v^2}^{-1/2}$.
We denote the so-called comoving derivative by $\td{}{t}$:
$\td{}{t}=\pd{}{t}+\v{v}\nabla$.

The thermodynamical quantities are the following: $\varepsilon$
is the energy density (including rest energy contribution), $p$
is the pressure, $w=\varepsilon+p$ is the enthalpy density, $T$
is the temperature, $\sigma$ is the entropy density. 
When there are some (conserved or non-conserved) charges present, we denote
them by $n_i$, and the corresponding chemical potentials by
$\mu_i$. All the above densities are defined in the  
local rest frame, so these densities are scalar quantities, while 
for example the entropy density in the laboratory frame is defined as
the 0-th component of the entropy current vector, $\sigma u^0$.

\subsection{The equations of relativistic hydrodynamics}

Assuming perfect fluid (that is, neglecting heat conductivity,
bulk and shear viscosity) the equations of relativistic
hydrodynamics are obtained by Landau's heuristic argumentation: we guess the form of the energy-momentum tensor
$T_{\mu\nu}$ in the local rest frame:
$T_{\mu\nu}=diag\z{\varepsilon,p,p,p}$, and from this we
have in arbitrary frame
\bl{e:Tmunu}
T_{\mu\nu}=wu_\mu u_\nu-pg_{\mu\nu} .
\ee
The equations then follow from the energy-momentum conservation
law $\partial_\nu T^{\mu\nu}=0$. Substituting \eq{e:Tmunu} and
projecting orthogonal and parallel to $u^\mu$, we obtain the
relativistic Euler equation and the energy conservation equation as
\bea
wu^\nu\partial_\nu u^\mu &=& \z{g^{\mu\rho}-u^\mu u^\rho}\partial_\rho p , \label{e:4deul} \\
w\partial_\mu u^\mu      &=& -u^\mu\partial_\mu\varepsilon . \label{e:4denergy}
\eea
The general form of the charge conservation equations (for one type of charge)
is as follows:
\bl{e:ccons}
\sum_i \mu_i \partial_\mu\z{n_iu^\mu}=0 ,
\ee
this is valid for the case of non-conserved charges, as their
chemical potentials vanish, for one conserved charge $n$
with non-vanishing chemical potential, as the conservation
law is $\partial_\mu\z{nu^\mu}=0$, and also for various
mixtures. (For instance, in case of baryonic or electric charge,
particles and antiparticles carry opposite charges, and they
chemical potentials are the same but of opposite sign, so
\eq{e:ccons} is valid also for this case.)

For latter usage we write down these equations also in a
three-dimensional notation. The relativistic Euler equation,
the energy conservation equation and the continuity equation
(for one conserved charge) are
\bea
\frac{w}{1-v^2}\td{\v{v}}{t} &=& -\z{\nabla p+\v{v}\pd{p}{t}} ,                    \label{e:3deul}    \\
\rec{w}\td{\varepsilon}{t} &=& -\z{\nabla\v{v}}-\rec{1-v^2}\td{}{t}\frac{v^2}{2} , \label{e:3denergy} \\
\td{}{t}\ln\frac{n}{\sqrt{1-v^2}} &=& -\z{\nabla\v{v}} . \label{e:3dcont}
\eea
The thermodynamical quantities obey general rules, which
follow directly from thermodynamics:
\bea
\m{d}\varepsilon &=& T\m{d}\sigma+\sum_i\mu_i\m{d}n_i , \label{e:GD1} \\
w                &=& T\sigma+\sum_i\mu_in_i ,           \label{e:GD2} \\
\m{d}p           &=& \sigma\m{d}T+\sum_in_i\m{d}\mu_i . \label{e:GD3}
\eea
The last equation is a consequence of the first two ones.
From \eqs{e:GD1}{e:GD3} and \eq{e:4denergy} one can derive
the entropy conservation equation:
\bl{e:scons}
\partial_\mu\z{\sigma u^\mu}=0 .
\ee
This is a consequence of the special choice of the
energy-momentum tensor. The perfectness of the fluid can
be defined alternatively by requiring that \eq{e:scons} holds.

In order to get a closed set of equations, we have to specify
the Equation of State (EoS) of the flowing matter. In the following we shall
investigate a case, which resembles to that of ideal gas, that is
\bl{e:ekappap}
\varepsilon=\kappa p .
\ee
The $\kappa$ constant is $3$ for a three-dimensional, ultra-relativistic
ideal gas, but now we retain the more general possibility of arbitrary
$\kappa$. With the choice of \eq{e:ekappap} the charge conservation
equation decouples from the Euler and the energy conservation equations.
That is, in order to find new solutions, one has to deal with $\v{v}$
and $p$ only, because \eqs{e:3deul}{e:3denergy} contains only
these variables, and charge densities occur in the calculation only
after one found proper solutions for $p\z{t,\v{r}}$ and $\v{v}\z{t,\v{r}}$.

An important property of this specific EoS is, that if one finds 
a solution with this EoS, then these solutions can be generalized
for a bag-model type of EoS in a straight-forward manner, 
as the following replacements leave the hydrodynamical equations invariant:
\bea
	\varepsilon & \rightarrow & \varepsilon+B, \\
	p & \rightarrow & p-B.
\eea
The value of the bag constant, $B$ cancels
from \eqs{e:3deul}{e:3denergy}, that is, one can introduce an arbitrary 
value of the bag constant. This does not change the time evolution of the
entropy density or the flow field as long as it is a constant,
independent of position and time.
Because this possibility is simple, we do not explicitly
write down the bag constant in the following calculations, but
emphasize, that this can be done without difficulty.

\subsection{Rindler coordinates}

We discuss below exact solutions of relativistic perfect fluid
hydrodynamics in 1+1 dimensions and spherical solutions in 1+d
dimensions as well (the 1+1 dimensional case is a special case
of this latter one). The number of spatial dimensions is denoted
by $d$. The notation $r$ stands for the $r_z$ spatial coordinate
in 1+1 dimensions, and for the radial coordinate in 1+d
dimensions. All quantities depend only on $t$ and $r$, and the
velocity field is radial. We use the well-known Rindler coordinates
$\tau$ and $\eta$ as independent variables, which naturally fit
to the Hwa-Bjorken solution.

These coordinates have different definitions for $t>r$ (we will
refer to this case as ,,inside the forward lightcone'') and for
$t<r$ (that is, outside the lightcone). We have
\bl{e:Rin}
t=\tau\cosh\eta \sp r=\tau\sinh\eta
\ee
inside the lightcone and
\bl{e:Rout}
t=\tau\sinh\eta \sp r=\tau\cosh\eta
\ee
outside the lightcone. 

In the next section we use these coordinates
in order to find new solutions of the hydrodynamical equations. We
investigate an ansatz for the flow and pressure field, which leads
to a new and interesting class of exact, explicit and simple solutions.
Among these solutions we find ones that overcome the shortcoming of
almost all presently known exact solutions: the lack of
acceleration.

\section{Deriving new solutions}

We parametrize the velocity with the $\Omega(\tau,\eta)$ rapidity as
$v=\tanh\Omega(\tau,\eta)$.
In this section we take $t>r$, that is we are inside the forward lightcone.
The Rindler variables $\tau$ and $\eta$ are defined by \eq{e:Rin}.
The Bjorken-solution is easily written down in terms of these
variables saying $\Omega=\eta$, and $\frac{\sigma}{\sigma_0}=\frac{\tau_0}{\tau}$.
Our new ansatz is the following generalization of the Bjorken flow:
\bl{e:lansatz}
\Omega=\lambda\eta \sp \lambda\in\mathbb{R} .
\ee
Here $\lambda$ is an (up to now) arbitrary constant, a  parameter of the
solution.
Substituting \r{e:lansatz} into the hydrodynamical equations (the
Euler and energy conservation equations with the $\varepsilon=\kappa p$ EoS)
after some calculations detailed in Appendix A we obtain that $p$ must
have the following form:
\bl{e:pgeneral}
p=p_0\z{\frac{\tau_0}{\tau}}^{-K\z{\kappa+1}}\rec{\cosh^{\frac{K+\lambda}{\lambda-1}\z{\kappa+1}}\z{(\lambda-1)\eta}} .
\ee
with $K$ being a constant. The value of $\lambda$ and $K$ are constrained
by the equation
\[
\kappa K+\z{d-1}\frac{\sinh\lambda\eta}{\sinh\eta}\cosh\z{(\lambda-1)\eta}+\mbox{}
\]
\bl{e:finaleq_}
\mbox{}+\lambda\z{\cosh^2\z{(\lambda-1)\eta}-\kappa\sinh^2\z{(\lambda-1)\eta}}=0 .
\ee
The detailed calculations which lead to these equations are found in Appendix A.

Let us emphasize, that \eq{e:finaleq_} is derived from the hydrodynamical equations
including the equations of state within lightcone, $|t| > |r|$, using Rindler coordinates.
Outside the lightcone, for $|t| < |r|$, this equation is modified to \eq{e:finaleq_out}, 
that governs the ``external'' hydrodynamical  solutions.

In the next subsection we summarize and investigate the possible solutions.

\section{New exact solutions}

There are 4 different sets of the parameters $\lambda$, $d$, $\kappa$ and $K$
which solve \eq{e:finaleq_}, as discussed below. 
A fifth solution in this class is the $\lambda=1$ particular case, 
which is known as the Hwa-Bjorken solution in 1+1 dimensions, and
whose multi-dimensional generalizations, 
often called as Buda-Lund type of relativistic solutions,
were discovered in the past years~\cite{Csorgo:2003rt,Csorgo:2003ry}.

In Appendix B, we present the proof that these five solutions 
are the only possible non-trivial solutions in this class 
(where $\Omega=\lambda\eta$.) In Appendix B, 
we also prove the  uniqueness  of these solutions in a broader
sense: If one allows $\lambda$ to be not only a constant but a proper-time
dependent function (i.e. if $\Omega=\lambda(\tau)\eta$), no
other exact solutions exist, except the five solutions 
a.) - e.) listed in Table~\ref{t:sol}. The properties of these solutions
are detailed below.

\subsection{Flow profile and pressure}

\begin{figure}
\includegraphics[height=240pt,angle=-90]{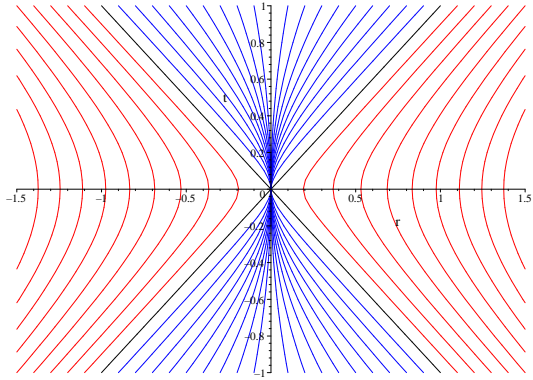}
\caption{\label{f:l2trajec} (Color online) Fluid trajectories of the $\lambda=2$ solution.
}
\end{figure}

In all cases, the velocity field and the pressure is expressed as
\bea
v&=&\tanh\lambda\eta , \label{e:vsol}\\
p&=&p_0\z{\frac{\tau_0}{\tau}}^{\lambda d\frac{\kappa+1}{\kappa}}\z{\cosh\frac{\eta}{2}}^{-(d-1)\phi_{\lambda}} . \label{e:psol}
\eea
Table~\ref{t:sol} shows the possible cases: every row represents one solution.
\begin{table}
\begin{tabular}{|c|c|c|c|c|}
  \hline
  Case & $\lambda$ &       $d$             & $\kappa$ &         $\phi_{\lambda}$            \\ \hline
  a.)  & $2$             & $\in\mathbb{R}$ & $d$              & $0$                         \\
  b.)  & $\rec{2}$       & $\in\mathbb{R}$ & $1$              & $\frac{\kappa + 1}{\kappa}$ \\
  c.)  & $\frac{3}{2}$   & $\in\mathbb{R}$ & $\frac{4d-1}{3}$ & $\frac{\kappa + 1}{\kappa}$ \\
  d.)  & $1$             & $\in\mathbb{R}$ & $\in\mathbb{R}$  & $0$                         \\
  e.)  & $\in\mathbb{R}$ & $1$             & $1$              & $0$                         \\  \hline
\end{tabular}
\caption{The new family of solutions.}\label{t:sol}
\end{table}
In what follows, we discuss the various cases separately.
\begin{itemize}
\item Case $a.)$:
In this case, $\lambda=2$, and the solution is obtained 
in arbitrary number of spatial
dimensions, ($d$ is arbitrary), but then $\kappa$ should be equal to $d$.
For instance, in three dimensions $\kappa$ should be $3$, which is the
EoS of an ultrarelativistic ideal gas or a photon gas. The corresponding
flow profile has a non-vanishing relativistic  acceleration: 
$u_\nu \partial^\nu u^\mu \ne 0$.
 We will see that this solution
in the same form solves the hydrodynamical equations outside the lightcone, 
too. If we write down the velocity field in
Minkowskian coordinates, it has the form both for $t<r$ and $t>r$
\bl{e:l2v_tr}
v=\frac{2tr}{t^2+r^2} ,
\ee
and we can solve the equation of motion for an individual particle, that is, we can
determine the flow trajectories $r(t)$. For a fluid element that is at position $r_0$ at
$t_0$ time-point, we have
\bl{e:l2trajec}
r(t)=\rec{a_0}(\sqrt{1+(a_0 t)^2}+1)
\ee
with $a_0=\frac{2r_0}{\left|r_0^2-t_0^2 \right|}$. These trajectories have
constant $a_0$ acceleration in the local rest frame. A possible interesting
application of this is mentioned briefly later. The trajectories
of this solution are shown in Fig.~\ref{f:l2trajec}.
It has a realistic equation of state, corresponding to a massless ultra-relativistic 
gas in $d$ dimensions. 

Note that we have found this solution first, using other methods.
This solution can also be found using the criteria that the pressure
depends only on proper-time and the solution should work in any $d$  
number of dimensions. These criteria select this case and the well known the $\lambda = 1$ 
Hwa-Bjorken solution, detailed as case $d.)$.

\item Case $b.)$: In this case $\lambda=\rec{2}$, the number of spatial dimensions is arbitrary, but
$\kappa=1$. The flow is accelerating. This case together with case $c.)$ was found first in $d=3$ dimensions
by T.~S.~Bir\'o~\cite{Biro:2007pr}, and we generalized it for arbitrary $d$. An important point is that
this case and case $d.)$ are the only ones where the pressure field can have explicit $\eta$-dependence, and
$p$ can be finite in $\eta$ (actually, it is finite if and only if $d\neq 1$).

\item Case $c.)$: This case is characterized with $\lambda=\frac{3}{2}$, $d$ is arbitrary, but
$\kappa=\frac{4d-1}{3}$, that is, for example for $d=1$ we get $\kappa=1$, and for $d=3$ it gives
$\kappa=\frac{11}{3}$. This case also was found in $d=3$ by T.~S.~Bir\'o first, and we
have generalized it. The pressure field is also finite in $\eta$ for $d\neq 1$.

\item Case $d.)$: In this case $\lambda=1$, $d$ and $\kappa$ can be arbitrary.
For $d=1$ this case is known as the Hwa-Bjorken solution, and for $d>1$ these
solutions were found in refs.~\cite{Csorgo:2003rt,Csorgo:2003ry}. The flow profile in the $d > 1$ case is called Hubble flow. Thus these solutions
were already known inside the light cone, 
but we quoted them for the sake of completeness.  
They are also the basis of new solutions, the the extensions of these solutions 
to the region outside the lightcone. These solutions are discussed in
the next subsection.  Inside the light-cone, this flow is not accelerating, 
$u_\nu \partial^\nu u^\mu = 0$.

\item Case $e.)$: In this case, the value of the parameter $\lambda$ 
is arbitrary, but $d$ \emph{and} $\kappa$ must be equal to $1$. 
The flow is accelerating, $u_\nu \partial^\nu u^\mu \ne 0$ if $\lambda\neq 1$.
The velocity field is pretty general, but the EoS is very particular.
Nevertheless, this solution can be considered as a ,,smooth extrapolation''
between the previous cases, 
although this kind of generalization --- the arbitrary
$\lambda$ case, which can be expected knowing the previous cases
 --- works only with this special choice of $d$ and $\kappa$. 
In the forthcoming we shall use this solution to calculate 
the rapidity distribution. We show in Appendix C, that
the width of the observable rapidity distribution is 
controlled by the parameter $\lambda$. Hence in principle,
the value of the parameter $\lambda$ of the hydrodynamical solution 
can be obtained from measurements.
\end{itemize}

\subsection{Outside the lightcone}

The hydrodynamical equations can also be rewritten in Rindler-coordinates outside
the lightcone (that is, for $|t|<|r|$). We investigate here how the solutions presented
in the previous subsection can be extended to this domain. The equations are found in
Appendix A, here we only go through the possibilities. The statements below can be directly
checked if one substitutes these expressions  into \eq{e:finaleq_out}
 in Appendix A,
which is derived from the hydrodynamical equations in the $|t| < |r|$ region.

\begin{itemize}
\item Case $a.)$: This solution can be extended outside the lightcone in a
straightforward manner. The properties are the same: it works in
arbitrary $d$ with $\kappa=d$, the velocity field has the same expression in
Minkowskian coordinates as inside the lightcone (see \eq{e:l2v_tr}), hence it is
also uniformly accelerating. Fig.~\ref{f:l2trajec} shows the trajectories
outside the lightcone, too. These trajectories are given by the same form,
\eq{e:l2trajec}, outside as well as inside the lightcone.

\item Case $b.)$ and $c.)$: These solutions can be extended to $t<r$ only in the $d=1$ case,
but these extensions are special cases of the extension of case $e.)$ discussed below.
These extensions do not depend explicitly on $\eta$. (Such a dependence was only present
for $t>r$ only if $d\neq 1$.)

\item Case $d.)$: An interesting new result is that there are solutions outside the lightcone
which can be viewed as the extensions of the $\lambda=1$, or Hwa-Bjorken  internal solutions.
The velocity and the pressure have the form
\bea
v&=&\tanh\eta=\frac{t}{r} , \label{e:l1outv} \\
p&=&p_0\z{\frac{\tau_0}{\tau}}^{\kappa+1}\cosh^{-\frac{d-1}{\kappa}\z{\kappa+1}}\eta .
\eea
Moreover, this extension is an accelerating flow. The equation of the trajectories is
\bl{e:l1trajec}
r(t)=\rec{a_0}(\sqrt{1+(a_0 t)^2}+1)
\ee
for $r_0>t_0$. 
These trajectories are also of constant $a_0$ acceleration,
as the a.) ($\lambda=2$) case. 
However, the expression of $a_0$ is different from that of case $a.)$,
we have now $a_0=\frac{r_0}{r_0^2-t_0^2}$. 

It is interesting to note at this point the following.
As both the $\lambda=2$ and
the $\lambda=1$ external solutions are uniformly accelerating (in local rest frame),
they both contain event horizons.
This property of the uniform acceleration was utilized recently by
Kharzeev and Tuchin~\cite{Kharzeev:2005iz} to describe thermalization
in heavy ion reactions via the Unruh effect.

\item Case $e.)$: The case when $\lambda$ is arbitrary and $d=\kappa=1$
can be extended to $t<r$ without difficulties, the outside solution is
also an accelerating one. The velocity and the pressure is given by
\bea
v&=&\tanh\lambda\eta, \\
p&=&p_0\z{\frac{\tau_0}{\tau}}^{2\lambda} ,
\eea
but remember, that the definition of $\tau$ and $\eta$ are given by \eq{e:Rout}, which
differs from the case of $|t|>|r|$. 
\end{itemize}

\subsection{Temperatures and densities}

The previously presented pressure and velocity fields solve \eqs{e:3deul}{e:3denergy}, the Euler and the
energy conservation equations. It is specific to the $\varepsilon=\kappa p$ EoS utilized
here, that the continuity equation can be treated separately. But if we investigate the
temperature and perhaps other charge densities, the relation between $\varepsilon$ and
$p$ does not fully determine the EoS. In this subsection we solve the continuity equations for some
cases.

Before doing this, we introduce the scaling variable $S\z{\tau,\eta}$ with the definition
that its comoving derivative vanishes, that is (for one-dimensional or spherical flows)
\bl{e:Sdef}
\td{S}{t}=\pd{S}{t}+v\pd{S}{r}=0 .
\ee
Every specific flow profile has a corresponding $S$ function, which can be determined by
solving \eq{e:Sdef}. Table~\ref{t:S} contains the expressions of $S$ for the investigated
class of solutions. (A remark: $S$ is not uniquely determined by \eq{e:Sdef}, if $S$ is a good scaling function,
then any function of $S$ is also a good scaling function. But we shall see, that this ambiguity does not really matter,
as it cancels from the observables.)

After this preparation, let us describe the solutions to the entropy (or in general, continuity type of)
equations.

\begin{table}
\begin{tabular}{|c||c|c|}
  \hline
  $S\z{\tau,\eta} $ & $\lambda\neq 1$ & $\lambda=1$ \\
\hline
\hline
  $t>r$ & $\z{\tau_0/\tau}^{\lambda-1}\sinh\z{(\lambda-1)\eta}$ & $\eta$ \\
  \hline
  $t<r$ & $\z{\tau_0/\tau}^{\lambda-1}\cosh\z{(\lambda-1)\eta}$ & $\tau/\tau_0$  \\
  \hline
\end{tabular}
  \caption{The $S\z{\tau,\eta}$ scaling functions inside and outside the light-cone.
Note, that the $\lambda = 1$ case has to be handled separately from the other cases.}\label{t:S}
\end{table}

First, if there are no charges present at all, then from the \r{e:GD2} Gibbs-Duhem relation we have
$T\sigma=\z{\kappa+1}p$, and it can be verified, that the solution of the entropy conservation ,
\eq{e:scons} is
\bea
\sigma &=& \sigma_0\z{\frac{p}{p_0}}^{\frac{\kappa}{\kappa+1}}\nu_\sigma(S) , \label{e:ssol} \\
T      &=& T_0\z{\frac{p}{p_0}}^{\frac{1}{\kappa+1}}\rec{\nu_\sigma(S)} \label{e:Tsol}
\eea
where $p\z{\tau,\eta}$ is given by \eq{e:psol},
and of course $\z{\kappa+1}p_0=T_0\sigma_0$. The function $\nu_\sigma(S)>0$ can be chosen arbitrarily.
This approximation (that is, the baryonic charges are negligibly small) is often used
for the description of high energy heavy ion collisions and of the quark-hadron transition of the early Universe.

Second, if we have one conserved charge $n$, and as an ideal gas EoS
we have $p=nT$, then the solution of the conservation equation is very similar:
\bea
n &=& n_0\z{\frac{p}{p_0}}^{\frac{\kappa}{\kappa+1}}\nu_n(S) , \\
T &=& T_0\z{\frac{p}{p_0}}^{\frac{1}{\kappa+1}}\rec{\nu_n(S)} .
\eea
The constants are chosen so that $p_0=n_0T_0$.

The case of more than one charges can be treated similarly: we have
$(\kappa+1)p=T\sigma+\sum_i n_i\mu_i$, and the solution is
\bea
\sigma &=& \sigma_0\z{\frac{p}{p_0}}^{\frac{\kappa}{\kappa+1}}\nu_\sigma (S) , \\
T      &=& T_0\z{\frac{p}{p_0}}^{\frac{1}{\kappa+1}}\rec{\nu_\sigma (S)} , \\
n_i    &=& n^{(i)}_0\z{\frac{p}{p_0}}^{\frac{\kappa}{\kappa+1}}\nu_i(S) , \\
\mu_i  &=& \mu^{(i)}_0\z{\frac{p}{p_0}}^{\frac{\kappa}{\kappa+1}}\mathcal{M}_i(S) ,
\eea
where the functions and the constants are constrained as
$\nu_i(S)\mathcal{M}_i(S)=1$ and $(\kappa+1)p_0=T_0\sigma_0+\sum_i n_0^{(i)}\mu_0^{(i)}$.
These forms are solutions to the hydrodynamical and continuity equations simultaneously.

Let us emphasize, that the scaling functions of the scaling variable $S$ are quite arbitrary,
they are only constrained by using a physical normalization, which implies $\nu_i(0) = \mathcal{M}_i(0) = 1$.
As we mentioned earlier, any function $S^\prime$ of the scaling variable $S$ is again a good scaling
variable, as long as $S^\prime(0) = 0$, because the same solutions are obtained 
when new scaling functions are chosen, that correspond to the new scaling variable
as $\nu_i^\prime(S^\prime(S)) = \nu_i(S)$. Thus it is possible to use the most natural form
of the scaling variable $S$, when describing these solutions.

\section{The rapidity distribution}

If we want to apply our solutions to the description of some physical phenomena,
we have to calculate the final state observables from these solutions. We have chosen
the rapidity distribution, and in order to be able to fit measured data, we
apply here our new solution described as case $e.)$ in the previous section.
Here the parameter $\lambda$ (which we can call ,,acceleration parameter'',
because it somehow influences the acceleration of the flow) can be arbitrary,
but the dimension is restricted to $d=1$, and what is more inconvenient,
$\kappa$ is also equal to $1$. But the $\lambda$ parameter can be fitted
to measured datasets, and this compensates the drawback of the specific EoS.

The detailed calculation of the rapidity distribution is presented in Appendix
C, here we briefly summarize the main steps. We use Boltzmann approximation,
and neglect baryonic charges and chemical potentials, so $T$ and $\sigma$ are given by
\eqs{e:ssol}{e:Tsol}. Although the solution is valid only for the EoS of massless
particles, we assume that at the freeze-out particles of rest mass $m$ appear
(e.g. pions with $m=140$MeV). We have chosen the freeze-out condition as
follows: the freeze-out hypersurface is pseudo-orthogonal to the four-velocity
field $u^\mu$, and the temperature at $\eta=0$ reaches a given $T_f$ value.
The equation of this hypersurface is
\bl{e:fout}
\z{\frac{\tau_f}{\tau}}^{\lambda-1}\cosh\z{(\lambda-1)\eta}=1 ,
\ee
where $\tau_f$ is the proper-time coordinate when the temperature reaches $T_f$ at $\eta=0$
(see Appendix C for details).

With a saddle-point integration in $\eta$, for $\lambda>1/2$, $m/T_f \gg 1$ and
$\nu_{\sigma}(s)=1$ we got
\bl{e:dndy-approx}
\td{n}{y}\approx\td{n}{y}\Big{|}_{y=0}
                       \cosh^{\pm\frac{\alpha}{2}-1}\z{\frac{y}{\alpha}}
                       e^{-\frac{m}{T_f}\sz{\cosh^\alpha\z{\frac{y}{\alpha}}-1}} ,
\ee
with $\alpha=\frac{2\lambda-1}{\lambda-1}$.
In an actual fit to data, the parameter $\alpha$ can be determined
more directly, than $\lambda$. From the value of the shape parameter $\alpha$, 
the parameter of the acceleration $\lambda$ can thus be determined from 
the inverse formula as
\bl{e:lambdaalpha}
\lambda = \frac{\alpha -1}{\alpha -2}.
\ee
This equation also indicates, that the flat rapidity distribution
corresponds to $\lambda \rightarrow 1 $ and $\alpha \rightarrow \infty$.

In the formula \eq{e:dndy-approx} and in the subsequent 
discussion of this distribution the upper sign
always means the true $1+1$ dimensional case (that is, not only the flow, but the produced
particles have only one degree of freedom), while the lower sign is for the case when
the $1+1$ dimensional solution is embedded in the 1+3 dimensional space-time,
and the produced particles have transverse degrees of freedom as well.
The embedding means that the fluid occupies infinite volume, and it is
homogeneous (it has no velocity, no pressure gradient components) in transverse directions.

\begin{figure}
\includegraphics[width=200pt]{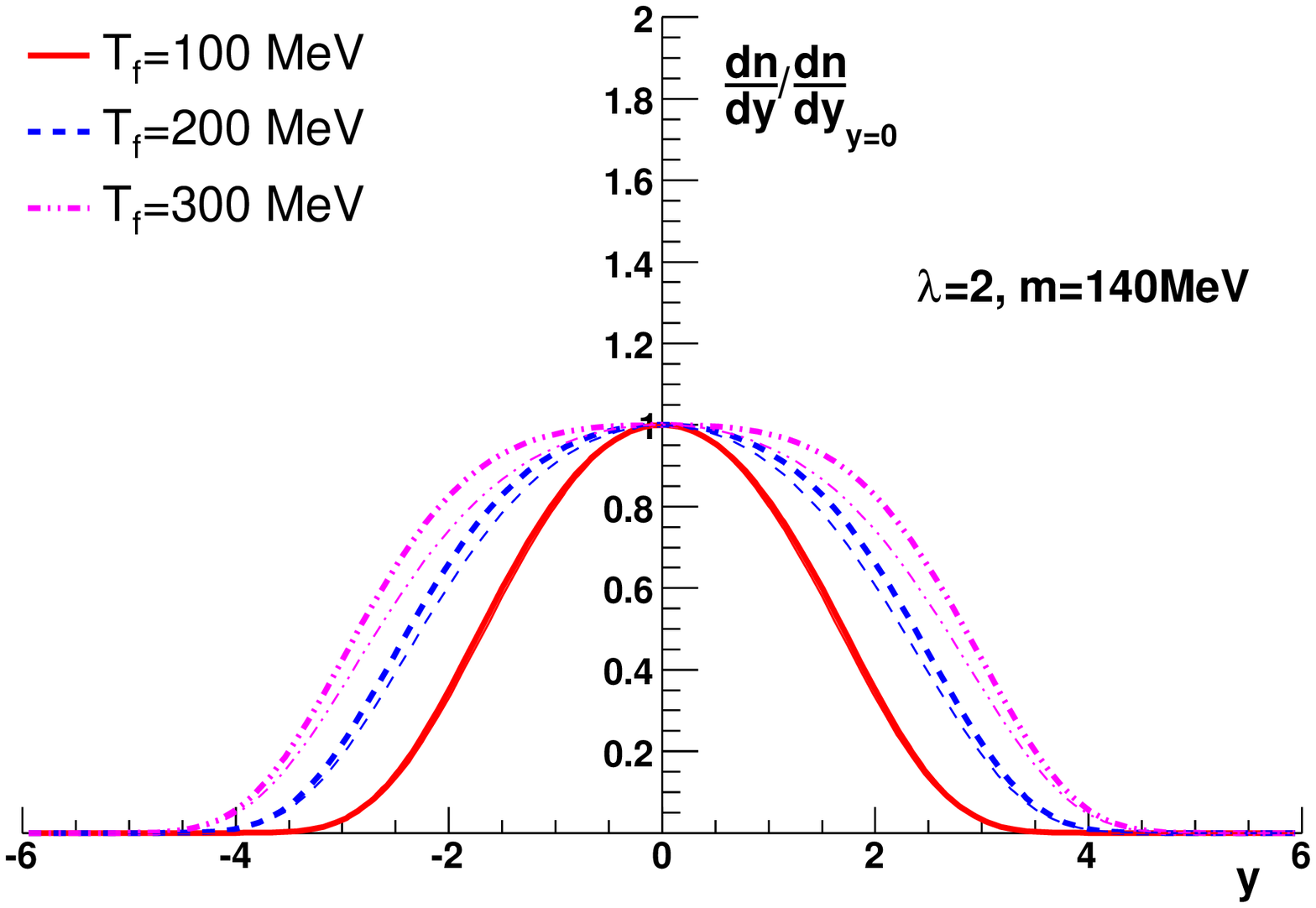}
\includegraphics[width=200pt]{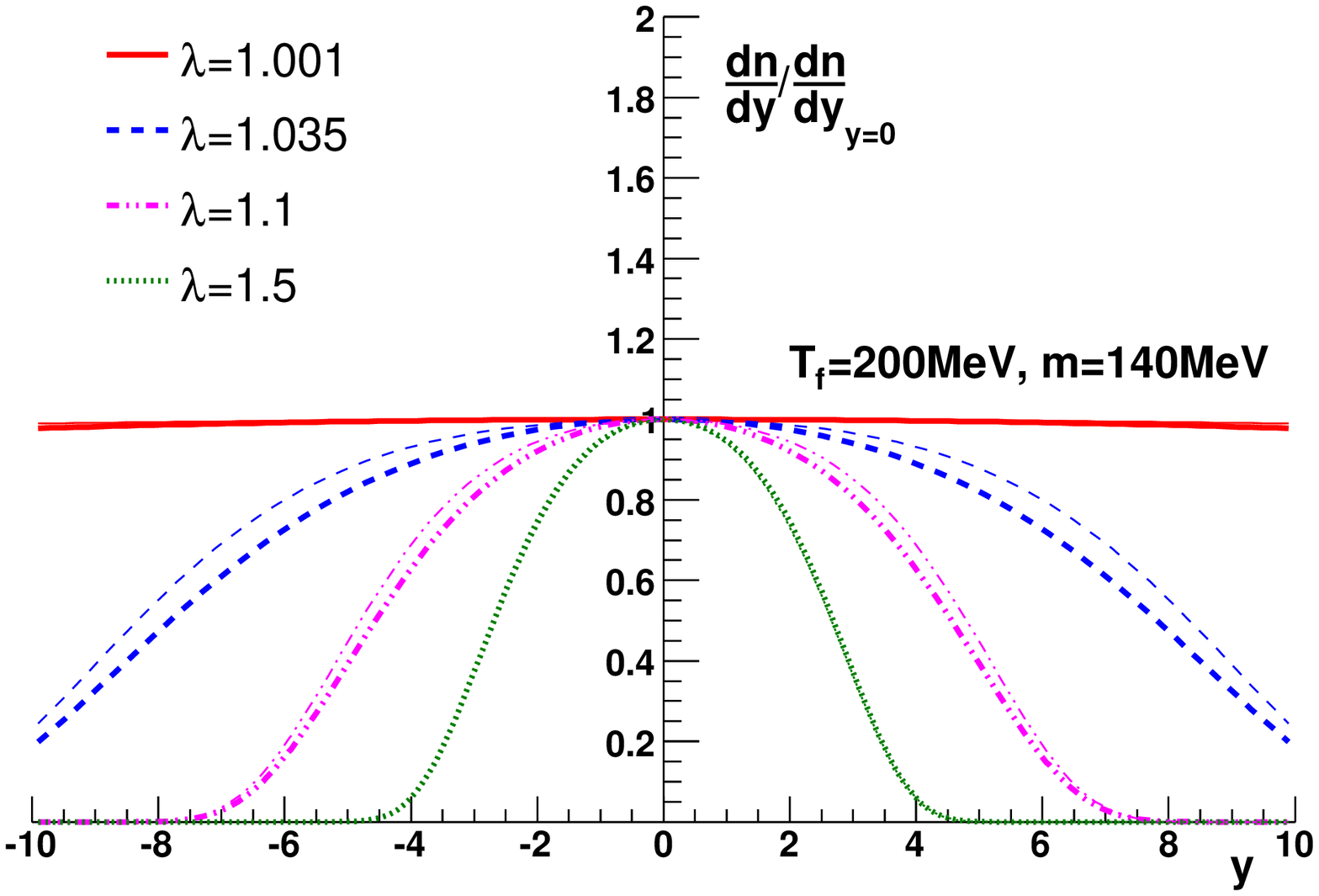}
\includegraphics[width=200pt]{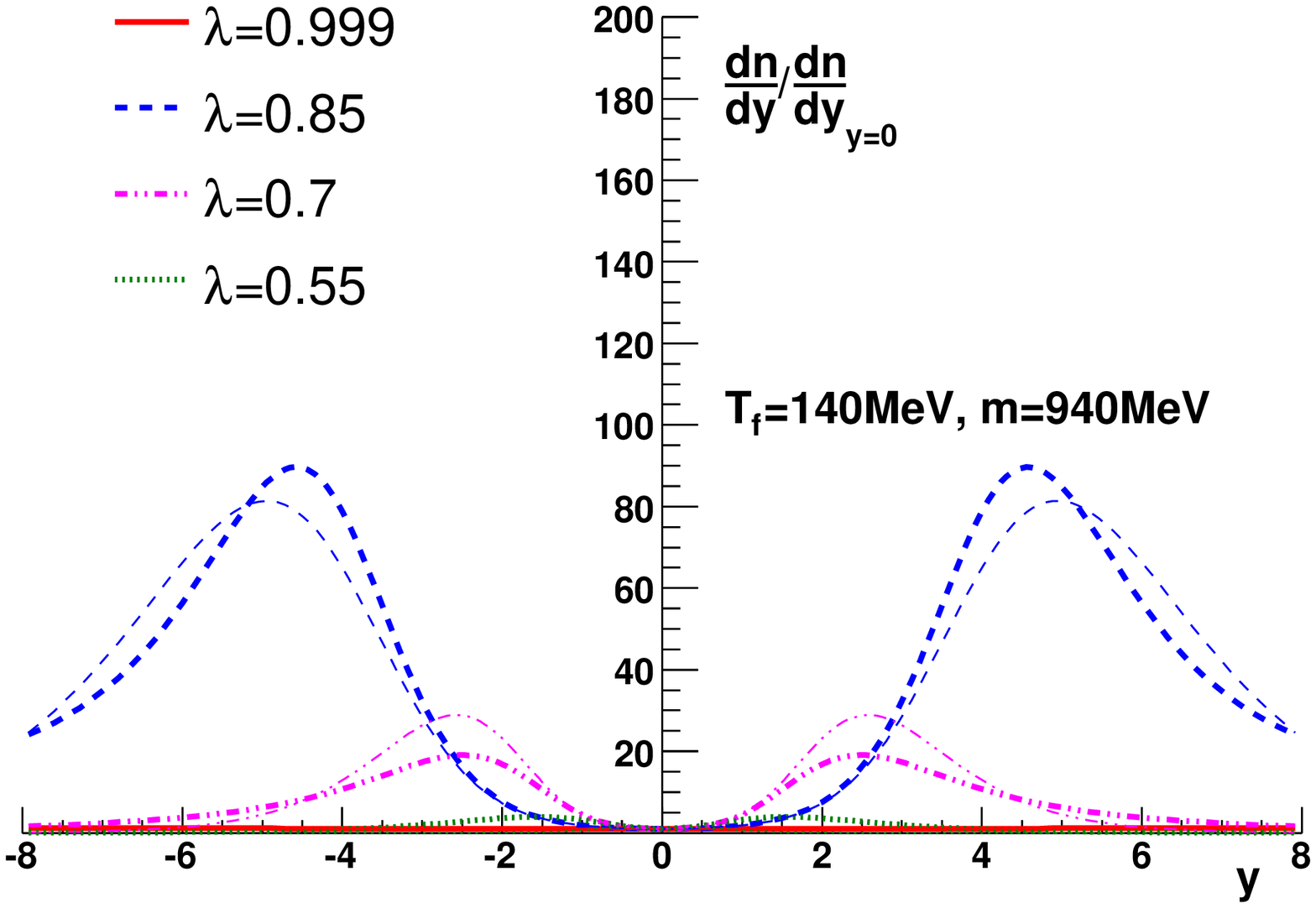}
\includegraphics[width=200pt]{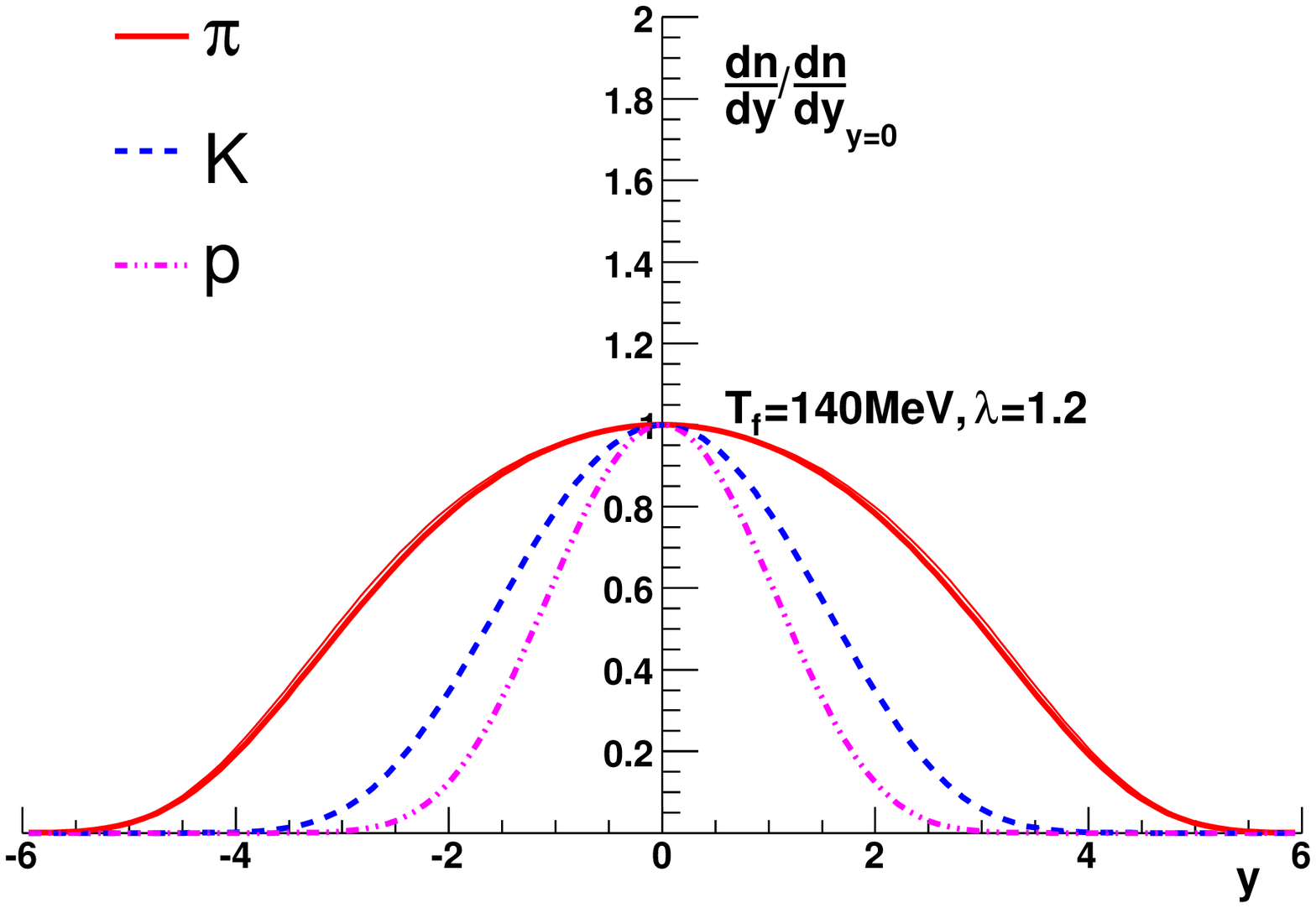}
\caption{\label{f:rapdist}(Color online) Normalized rapidity
distributions from the new solutions in 1+1 dimensions for various
$\lambda$, $T_f$ and $m$ values. Thick lines
show the result of numerical integration, thin lines
the analytic approximation from \eq{e:dndy-approx}. For $\lambda>1$
and not too big $T_f$ it can be used within about 10 \% error. }
\end{figure}

The Gaussian ,,width'' of the eq. ~\eq{e:dndy-approx} 
distribution (obtained from the second derivative
at $y=0$) is
\bl{Delta}
    \Delta y^2 = \frac{\alpha }{m/T_f \mp 1/2 + 1/\alpha} .
\ee
We see from the expression of $\Delta y^2$ that it changes sign if $\lambda$ (and thus $\alpha$)
varies. It has two roots in $\lambda$: $\Delta y^2=0$ if $\lambda = 1$,
or $\lambda = \rec{2}\z{1+\frac{T_f}{2m\mp T_f+T_f}}$. If $\lambda=1$ (this is the Hwa-Bjorken
case), the distribution is entirely flat, and if $\lambda$ equals the other root, it is not flat
but has only higher order non-vanishing derivatives at $y=0$. In other cases, the sign of
$\Delta y^2$ determines the behavior of the rapidity distribution (around $y=0$): it has a
minimum at $y=0$, if $\Delta y^2<0$, and it has a maximum (that is, it is approximately
Gaussian) if $\Delta y^2>0$. The typical cases are plotted in Fig.~\ref{f:rapdist}.
This figure is just an illustration, the parameter values on this figure
are not realistic ones, they were chosen in order to make the behavior of the
distribution transparent in the different cases.

\section{Applications}

In the previous section we have calculated the rapidity distribution
from our new solution. We are now able to fit real rapidity measurements
with our calculation, and we can extract the $\lambda$ parameter
from the data. (Fig.~\ref{f:rapdist} shows that the analytic
approximation for not too high temperatures has less than ~10\%
error, so it can be used reliably.) We can also see on
Fig.~\ref{f:rapdist} that in the $\lambda\to 1$ limiting case
the distribution approximates the Bjorken-like flat one, and the
more $\lambda$ differs from $1$, the more the distribution deviates
from the flat shape. That is, the acceleration effects make the
distribution a finite one, and since our calculation takes
acceleration into account, it provides a realistic description
of the real distributions, which have finite widths. We present
two applications: the improvement of the Bjorken energy density
estimation, and an advanced reaction life-time estimation.

\subsection{Energy density estimation}

We follow Bjorken's method~\cite{Bjorken:1982qr} and deviate from
it at the point when acceleration effects (which are neglected by
the Bjorken estimate) become important. Let us focus on a thin
transverse piece of the produced matter at mid-rapidity, illustrated
by Fig. 2 of ref.~\cite{Bjorken:1982qr}. The radius $R$ of this slab
is estimated by the radius of the colliding hadrons or nuclei:
$R=1.18A^{1/3}$fm. Its volume is $dV=(R^2\pi)\tau\m{d}\eta$, where
$\tau$ is the proper time of observation and $\m{d}\eta$ is the
space-time rapidity element corresponding to the slab. The energy content
in this slab is $dE =\langle m_t\rangle dn$, where $\langle m_t\rangle$
is the average transverse mass at mid-rapidity, so similarly
to Bjorken, the initial energy density is
\bl{e:Bjorken}
    \varepsilon_0 = \frac{\langle m_t\rangle}{(R^2 \pi)\tau_0}\frac{dn}{d\eta_0} .
\ee
Here $\tau_0$ is the proper-time of thermalization. (Bjorken's
estimate was $\tau_0\approx 1$fm.) For accelerationless,
boost-invariant Hwa-Bjorken flows $\eta_0=\eta_f=y$, however,
for our accelerating solution we have to apply a correction
factor of $\pd{y}{\eta_f}\pd{\eta_f}{\eta_0}$. From our
$\lambda\in\mathbb{R}$ solutions the shift of the point of
maximum emittiviy is $\pd{y}{\eta_f}=\z{2\lambda-1}$,
while the volume element change is
$\pd{\eta_f}{\eta_0}=\z{\tau_f/\tau_0}^{\lambda-1}$.
These two factors contain the acceleration effects on the
energy density estimation: the first one is the effect of
work done by the pressure, and the second one characterizes
how acceleration influences the expansion of the initial
volume element. For an accelerating flow both factors
should be greater than $1$, which is indeed the case if $\lambda>1$.

\begin{figure}
\includegraphics[height=240pt,angle=-90]{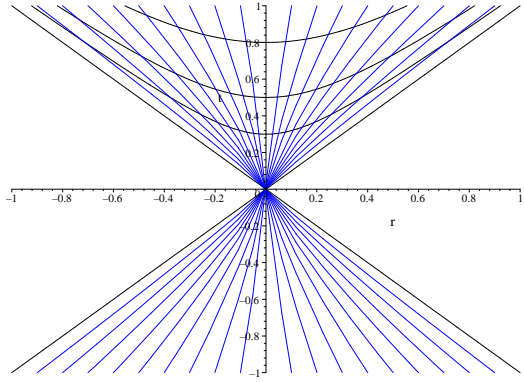}
\caption{\label{f:freezeout} (Color online) Fluid trajectories of the $\lambda=1.2$ solution,
which fits to BRAHMS $\td{n}{y}$ data. We illustrated the possible hypersurfaces (the hypersurfaces
pseudo-orthogonal to the flow trajectories), among these ones there is the proper freeze-out
hypersurface, selected by the criterion $T\z{\eta=0}=T_f$.
}
\end{figure}

\begin{figure}%[htb]
\includegraphics[width=200pt]{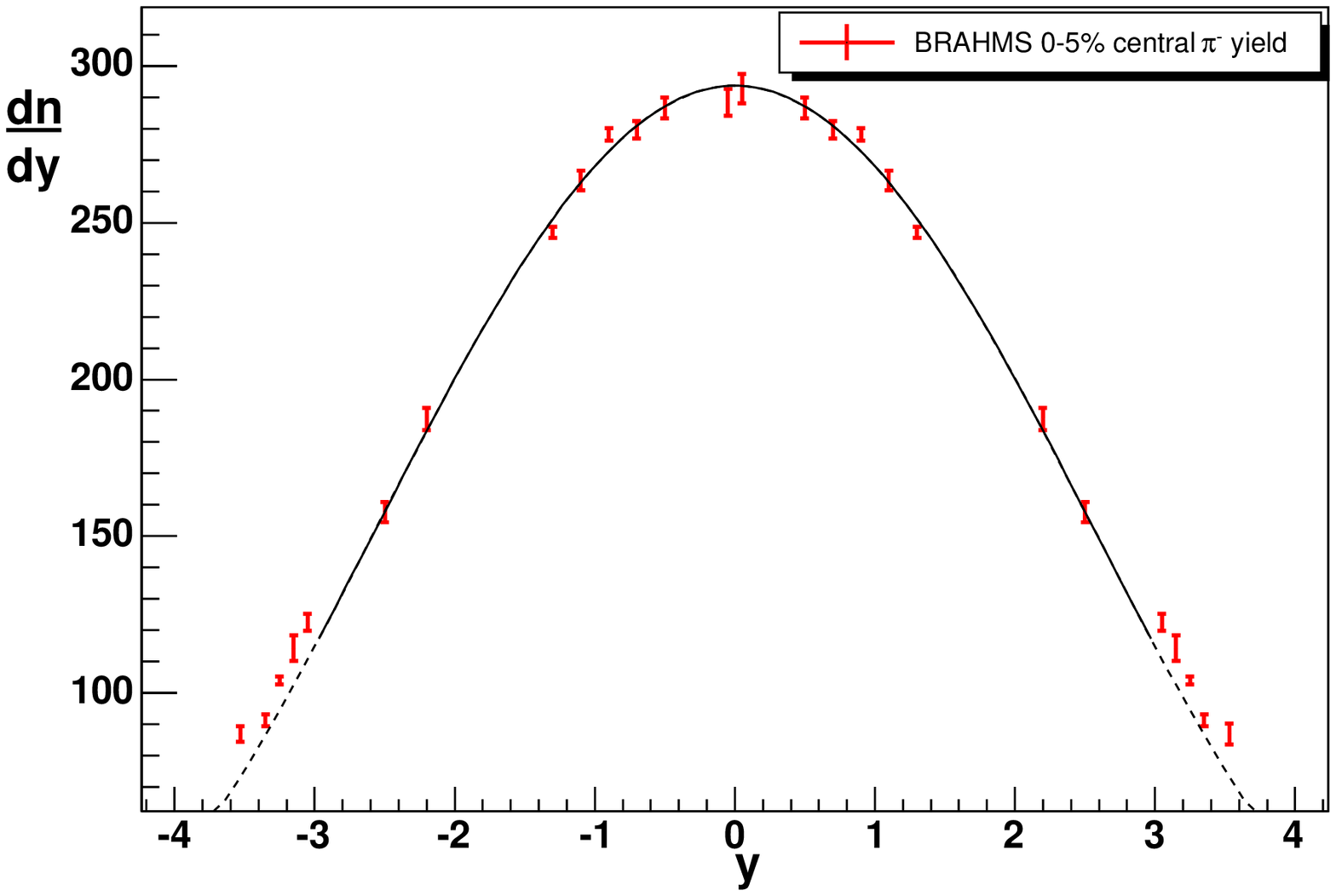}
\includegraphics[width=200pt]{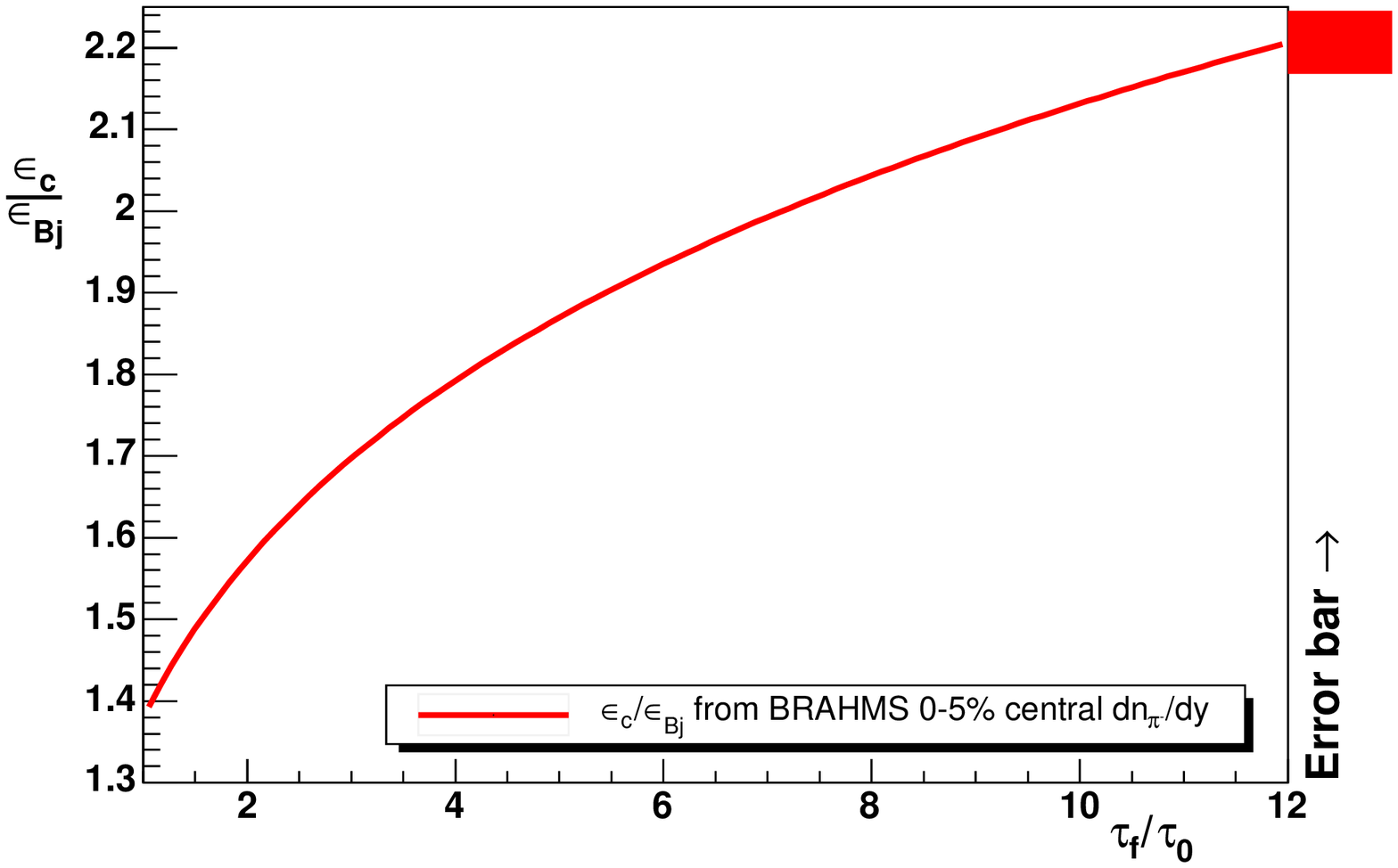}
\caption{\label{f:estim}(Color online)
Panel (a):
$dn/dy$ data of negative pions, as measured by
the BRAHMS collaboration~\cite{Bearden:2004yx} in central (0-5\%)
Au+Au collisions at $\sqrt{s_{NN}}=200$ GeV, 
fitted with \eq{e:dndy-approx} (1+3 dimensional case, lower signs).
The fit range was $-3 < y < 3$, to exclude
target and projectile rapidity region, CL = 0.6 \%. The parameters of the fit
are summarized in Table~\ref{table:2}.
Panel (b): $\varepsilon_c/\varepsilon_{Bj}$ ratio as a function
of $\tau_f/\tau_0$.
}
\end{figure}

Thus the initial energy density $\varepsilon_0$ can be accessed by an
advanced estimation $\varepsilon_c$ as
\bl{e:ncscs}
\frac{\varepsilon_c}{\varepsilon_{Bj}}=\z{2\lambda-1}\z{\frac{\tau_f}{\tau_0}}^{\lambda-1}
\sp \varepsilon_{Bj}=\frac{\langle m_t\rangle}{(R^2\pi)\tau_0}\frac{dn}{dy} .
\ee
Here $\varepsilon_{Bj}$ is the Bjorken estimation, which is recovered
if $\td{n}{y}$ is flat (i.e. $\lambda=1$), but if $\lambda>1$,
$\varepsilon_0$ is \emph{under-estimated} by the Bjorken formula.
Fig.~\ref{f:estim} shows our fits to BRAHMS $dn/dy$
data~\cite{Bearden:2004yx}. From these fits we have found 
$\lambda=1.18\pm 0.01$. On Fig.~\ref{f:freezeout} we plotted the fluid trajectories
and illustrated the ensemble of possible freeze-out hypersurfaces, as given by \eq{e:fout}
inside the lightcone for this specific $\lambda$ value. For illustration we also
calculated the rapidity distribution for $\lambda=1.18$: the bottom panel on
Fig.~\ref{f:rapdist} shows it with this $\lambda$ value for pions, kaons and protons. 

Using the Bjorken estimate of $\varepsilon_{Bj} = 5$ GeV/fm$^3$ as given in
ref.~\cite{BRAHMS-White}, and $\tau_f/\tau_0=8\pm 2$ fm/c, we find
an initial energy density of $\varepsilon_c = (2.0\pm
0.1)\varepsilon_{Bj}=10.0\pm 0.5$ GeV/fm$^3$. If the evolution
deviates from a $1+1$ dimensional perfect flow, then our
estimation is only a lower limit for the initial energy density.

\begin{table}
\begin{tabular}{|c|c|}
  \hline \hline
  $\alpha$  &  7.4 $ \pm $ 0.13 \\ 
  $\frac{\mathrm{d}n}{{\mathrm{d}y}}|_{y=0}$  &  294 $ \pm $ 1   \\ \hline
  $\chi^2$/NDF & 30.6/14  \\ 
  CL &  0.6 \% \\ \hline
  $T_f$ (MeV)  & 200 \mbox{\rm (fixed)}   \\ 
  $m$ (MeV)  & 140 \mbox{\rm (fixed)}   \\ 
  $\lambda$ &  1.18 $\pm $ 0.01 \mbox{\rm (derived)} \\ \hline\hline
\end{tabular}
\caption{The parameters of the best  fit 
 with \eq{e:dndy-approx} 
(1+3 dimensional case, lower signs) 
to $dn/dy$ data of negative pions, as measured by
the BRAHMS collaboration~\cite{Bearden:2004yx} in central (0-5\%)
Au+Au collisions at $\sqrt{s_{NN}}=200$ GeV.
The free fit parameters were $\alpha$ and 
$\frac{\mathrm{d}n}{{\mathrm{d}y}}|_{y=0} $ . 
The fit range was $ - 3 < y < 3$ . 
The confidence level of the fit, CL was calculated from $\chi^2/NDF$ .
The temperature parameter $T_f$ was fixed to 200 MeV, corresponding to
 the slope parameter of the single particle spectra, as in 
the fitted 1+3 dimensional case a 1+1 dimensional 
exact hydro solution was embedded to a 1+3 dimensional space.
For the pion mass, a fixed 140 MeV value was used.
The parameter of the acceleration, 
$\lambda$ was calculated from the fitted parameter $\alpha$ 
using \eq{e:lambdaalpha}.
}
\label{table:2}
\end{table}

\subsection{Life-time estimation}

For a Hwa-Bjorken type of accelerationless, coasting  longitudinal flow,
Sinyukov and Makhlin~\cite{Makhlin:1987gm} determined the longitudinal
length of homogeneity as
\bl{e:SM_Rlong}
R_{long} = \sqrt{\frac{T_f}{m_t}}\tau_{Bj} .
\ee
Here $m_t$ is the transverse mass and $\tau_{Bj}$ is the (Bjorken)
freeze-out time. This result provides a means to determine the life-time
of the reaction: it can be simply identified with $\tau_{Bj}$.
However, if the flow is accelerating, the estimated origin of the
trajectories is shifted back in proper-time, so the life-time of the
reaction is under-estimated by $\tau_{Bj}$. (This was pointed out
also in refs.~\cite{Wiedemann:1999ev,Csanad:2004cj,Renk:2003gn,Pratt:2006jj}.)
From our solutions we have (for a broad but finite rapidity distribution,
where the saddle-point approximation is valid):
\bl{Rlong-c}
R_{long}=\sqrt{\frac{T_f}{m_t}}\frac{\tau_c}{\lambda}
    \quad\Rightarrow\quad
    \tau_{c} = \lambda \tau_{Bj} .
\ee
So the new estimation of the life-time, $\tau_c$ contains a $\lambda$
multiplication factor. As mentioned in the previous subsection, BRAHMS
rapidity distributions in Fig.~\ref{f:estim} yield $\lambda=1.18\pm 0.01$,
so they imply a 18 $\pm$1 \% increase in the estimated life-time of the
reaction.

\section{Summary}

In this paper, we presented new simple, explicit analytic solutions to
the relativistic hydrodynamical equations. These new solutions are characterized  by a
non-vanishing relativistic acceleration, which is a significant development compared
to the already known solutions. Many of these new solutions work not only
inside the lightcone, but outside of it as well. As a by-product, we
have demonstrated that the well-known Hwa-Bjorken solution also has such
an extension, and this extension turns out to have uniformly accelerating
trajectories.

We have calculated the rapidity distribution of the particles
from our new solutions numerically, and presented a simple and reliable analytic
approximation as well. These distributions have finite widths, hence
they can be used to fit real experimental rapidity distribution data.
We have fitted BRAHMS data, and obtained a value of $\lambda=1.18\pm
0.01$ for the acceleration parameter of the solution.

The famous Bjorken estimate of the initial energy density could be
refined, taking acceleration effects into account. Using the Bjorken
estimate given in the BRAHMS White Paper~\cite{BRAHMS-White}
($\varepsilon_{Bj} = 5$ GeV/fm$^3$) and $\tau_f/\tau_0=8\pm 2$ fm/c,
we find an initial energy density of $\varepsilon_c = (2.0\pm
0.1)\varepsilon_{Bj}=10.0\pm 0.5$ GeV/fm$^3$, which is a significant
increase. We have also given an improved estimation for the life-time
of the reaction: a 18$\pm$ 1 \% increase is due to the presence of
acceleration, based on our model of the initial accelerating period.

\section{Acknowledgements}

It is our pleasure to  thank to A. Bialas, T. S. Bir\'o, W. Florkowski,
M. Gyulassy, D. Kharzeev, Yu. Karpenko, S. Pratt, Yu. M. Sinyukov,
H. St\"ocker, K. Tuchin and W. A. Zajc for inspiring discussions.
T. Cs. would like to thank Y. Hama and T. Kodama for inspiring
discussions and an invitation at an early stage of this study.
This work was supported by the US-Hungarian Fulbright Foundation,
the Hungarian OTKA grants T038406 and T049466,
the FAPESP grants 99/09113-3 00/04422-7 and 02/11344-8 of S\~ao Paulo, Brazil,
the KBN-OM Hungarian-Polish grant PL-9/03
and the NATO PST.CLG.980086 grant.

Note that during the process of finalizing this manuscript,
we have encountered two interesting, new family of relativistic hydrodynamical solutions,
in ref.~\cite{Bialas:2007iu} and also in ref.~\cite{Borshch:2007es},
both of which look to be very interesting and deserve more detailed, further 
investigations.
 
%%%%%%%%%%%%%%%%%%%%%%%%%%%%%%%%%%%%%%%%%%%%%%%%%%%%%%%%%%%%%%%%%%%%%%%%%%%%%%%%%%%%%%%%%%

\appendix
\section{Appendix A: Hydrodynamical equations in Rindler coordinates}

In this Appendix we present the derivation of \eqs{e:pgeneral}{e:finaleq_} in more detail. First we
rewrite the hydrodynamical equations in terms of the Rindler variables, then investigate
the assumption for the velocity field made in \eq{e:lansatz}. In this Appendix, as in the
body of the paper, $\Omega$ stands for the rapidity of the flow, $v=\tanh\Omega$, and here we
use the notation $\td{}{r}=\pd{}{r}+v\pd{}{t}$, this appears on the r.h.s. of the Euler equation.
We treat the two different domains of the variables --- inside and outside the
lightcone --- carefully, because the Rindler coordinates have different definitions.
We shall see that the $\Omega=\eta$ case is a particular one. (We refer to it as
Hwa-Bjorken case, but emphasize that for $d>1$ such solutions were found only recently.) So
we investigate it in a separate subsection, in the following two subsections --- where we
derive the general equations inside and outside the lightcone --- we assume $\lambda\neq 1$.

\subsection{Inside the forward lightcone}

First, we deal with the $t>r$, $t>0$ case. The definition of the Rindler coordinates,
the inverse formulas and the derivation rules are as follows:
\bea
t=\tau\cosh\eta &\sp& r=\tau\sinh\eta , \\
\tau=\sqrt{t^2-r^2} &\sp& \eta=\m{arctanh}\frac{r}{t} ,
\eea
\bea
\pd{}{t}&=&\cosh\eta\pd{}{\tau}-\frac{\sinh\eta}{\tau}\pd{}{\eta}, \\
\pd{}{r}&=&-\sinh\eta\pd{}{\tau}+\frac{\cosh\eta}{\tau}\pd{}{\eta} , \\
\td{}{t}&=&\frac{\cosh\z{\Omega-\eta}}{\cosh\Omega}\pd{}{\tau}+\frac{\sinh\z{\Omega-\eta}}{\tau\cosh\Omega}\pd{}{\eta} ,\\
\td{}{r}&=&\frac{\sinh\z{\Omega-\eta}}{\cosh\Omega}\pd{}{\tau}+\frac{\cosh\z{\Omega-\eta}}{\tau\cosh\Omega}\pd{}{\eta} .
\eea
The domain of the variables is $-\infty<\eta<+\infty$, and $0\le\tau<\infty$.
We introduce the notation $Q=\rec{\kappa+1}\ln p$. (The convenience of this notation
is due to the special EoS we utilize here, the $\varepsilon=\kappa p$ one.)
Rewriting and rearranging the Euler and energy conservation equations we obtain the following:
\bl{e:eul_Rin}
\pd{\Omega}{\tau}+\rec{\tau}\pd{Q}{\eta}+\tanh\z{\Omega-\eta}\z{\pd{Q}{\tau}+\rec{\tau}\pd{\Omega}{\eta}}=0 ,
\ee
\[
\kappa\pd{Q}{\tau}+\rec{\tau}\pd{\Omega}{\eta}+\frac{d-1}{\cosh\z{\Omega-\eta}}\frac{\sinh\Omega}{\tau\sinh\eta}+\mbox{}
\]
\bl{e:energy_Rin}
\mbox{}+\tanh\z{\Omega-\eta}\z{\pd{\Omega}{\tau}+\frac{\kappa}{\tau}\pd{Q}{\eta}}=0 .
\ee
Here we used the expression for the divergence of the velocity field:
$\z{\nabla\v{v}}=\pd{v}{r}+\frac{d-1}{r}v$.

Our special assumption was that $\Omega=\lambda\eta$, so substituting $\pd{Q}{\eta}$
from \eq{e:eul_Rin} into \r{e:energy_Rin} we have
\bl{e:Qeq}
\tanh\z{(\lambda-1)\eta}\z{\tau\pd{Q}{\tau}+\lambda}+\pd{Q}{\eta}=0 ,
\ee
\[
\kappa\tau\pd{Q}{\tau}+\z{d-1}\frac{\sinh\lambda\eta}{\sinh\eta}\cosh\z{(\lambda-1)\eta}+\mbox{}
\]
\bl{e:lambdaeq}
\mbox{}+\lambda\z{\cosh^2\z{(\lambda-1)\eta}-\kappa\sinh^2\z{(\lambda-1)\eta}}=0 .
\ee
From \eq{e:lambdaeq} we see that $\tau\pd{Q}{\tau}$ must be $\tau$-independent, that is
$Q$ must have the form $Q=H\z{\eta}+K\z{\eta}\ln\tau$. Then from \eq{e:Qeq} we have that
$K\z{\eta}$ is also constant. Again with \eq{e:lambdaeq} we have then
\bl{}
Q=K\ln\tau+H\z{\eta} ,
\ee
where the $H\z{\eta}$ function is determined by \eq{e:Qeq}. The solution is
easily obtained as
\bl{}
H\z{\eta}=-\frac{K+\lambda}{\lambda-1}\ln\cosh\z{(\lambda-1)\eta}+const.
\ee
for $\lambda\neq 1$, so $p$ has the form
\bl{}
p=p_0\z{\frac{\tau_0}{\tau}}^{-K\z{\kappa+1}}\rec{\cosh^{\frac{K+\lambda}{\lambda-1}\z{\kappa+1}}\z{(\lambda-1)\eta}} .
\ee
The integration constant was absorbed into the definitions of the $p_0$
and $\tau_0$ initial values. For the $K$ constant we have from \eq{e:lambdaeq}
\[
\kappa K+\z{d-1}\frac{\sinh\lambda\eta}{\sinh\eta}\cosh\z{(\lambda-1)\eta}+\mbox{}
\]
\bl{e:finaleq}
\mbox{}+\lambda\z{\cosh^2\z{(\lambda-1)\eta}-\kappa\sinh^2\z{(\lambda-1)\eta}}=0 .
\ee
Essentially this is \eq{e:finaleq_}, so what we need is to find such $\lambda$
and $K$ constants, which solve this equation, then we have the pressure and the
velocity field. The solutions were found, they are presented in the body of the paper.

\subsection{Outside the forward lightcone}

In this case the calculation goes almost in the same way as inside the
lightcone. The definitions and the derivation rules are:
\bea
t=\tau\sinh\eta &\sp& r=\tau\cosh\eta , \\
\tau=\sqrt{r^2-t^2} &\sp& \eta=\m{arctanh}\frac{t}{r} ,
\eea
\bea
\pd{}{t}&=&-\sinh\eta\pd{}{\tau}+\frac{\cosh\eta}{\tau}\pd{}{\eta} , \\
\pd{}{r}&=&\cosh\eta\pd{}{\tau}-\frac{\sinh\eta}{\tau}\pd{}{\eta}  , \\
\td{}{t}&=&\frac{\sinh\z{\Omega-\eta}}{\cosh\Omega}\pd{}{\tau}+\frac{\cosh\z{\Omega-\eta}}{\tau\cosh\Omega}\pd{}{\eta} , \\
\td{}{r}&=&\frac{\cosh\z{\Omega-\eta}}{\cosh\Omega}\pd{}{\tau}+\frac{\sinh\z{\Omega-\eta}}{\tau\cosh\Omega}\pd{}{\eta} .
\eea
We have in this case for the hydrodynamical equations
\bl{e:eul_Rout}
\pd{Q}{\tau}+\rec{\tau}\pd{\Omega}{\eta}+\tanh\z{\Omega-\eta}\z{\pd{\Omega}{\tau}+\rec{\tau}\pd{Q}{\eta}}=0 ,
\ee
\[
\pd{\Omega}{\tau}+\frac{\kappa}{\tau}\pd{Q}{\eta}+\frac{d-1}{\cosh\z{\Omega-\eta}}\frac{\sinh\Omega}{\tau\cosh\eta}+\mbox{}
\]
\bl{e:energy_Rout}
\mbox{}+\tanh\z{\Omega-\eta}\z{\kappa\pd{Q}{\tau}+\rec{\tau}\pd{\Omega}{\eta}}=0 .
\ee
If we now assume $\Omega=\lambda\eta$, then going through the same steps as inside
the lightcone we have (for $\lambda\neq 1$)
\bl{}
p=p_0\z{\frac{\tau_0}{\tau}}^{-K\z{\kappa+1}}\sinh^{-\frac{K+\lambda}{\lambda-1}(\kappa+1)}\z{(\lambda-1)\eta} ,
\ee
and the equation which constrains $K$ and $\lambda$ is
\[
\frac{\kappa\z{K+\lambda}}{\sinh\z{(\lambda-1)\eta}}-\z{d-1}\frac{\sinh\lambda\eta}{\cosh\eta}+
\]
\bl{e:finaleq_out}
+\z{\kappa-1}\lambda\sinh\z{(\lambda-1)\eta}=0 .
\ee
This equation is analogous to \eq{e:finaleq}, but it is slightly different. We used this equation to investigate the solutions outside the lightcone.

\subsection{The Hwa-Bjorken case}

The $\lambda=1$ case is the Hwa-Bjorken solution in 1+1 dimensions, and the Buda-Lund
case (refs.~\cite{Csorgo:2003rt,Csorgo:2003ry}) in 1+d dimensions. which is simply
obtained from \eqs{e:eul_Rin}{e:energy_Rin}.
We have
\bl{}
\pd{Q}{\eta}=0 \sp \kappa\pd{Q}{\tau}=-\frac{d}{\tau} ,
\ee
and the solution is
\bl{}
\Omega=\eta \sp p=p_0\z{\frac{\tau_0}{\tau}}^{\frac{d}{\kappa}\z{\kappa+1}} .
\ee
What is perhaps more interesting, is that this solution can be extended to outside the lightcone,
as we see it from \eqs{e:eul_Rout}{e:energy_Rout}, if we substitute $\lambda=1$:
\bl{}
\pd{Q}{\tau}+\rec{\tau}=0 \sp \pd{Q}{\eta}=-\frac{d-1}{\kappa}\tanh\eta ,
\ee
so the ,,outside Hwa-Bjorken-Buda-Lund solution'' is
\bl{}
\Omega=\eta \sp
p=p_0\z{\frac{\tau_0}{\tau}}^{\z{\kappa+1}}\cosh^{-\frac{d-1}{\kappa}\z{\kappa+1}}\eta .
\ee
This is an accelerating solution, its properties are discussed in the body of the paper.

%%%%%%%%%%%%%%%%%%%%%%%%%%%%%%%%%%%%%%%%%%%%%%%%%%%%%%%%%%%%%%%%%%%%%%%%%%%%%%%%%%%%%%%

\appendix
\section{Appendix B: Taylor expansion and uniqueness}

In this Appendix we sketch the proof of uniqueness of the presented solutions.
We deal only with the $t>r$ case. We investigate a less restrictive assumption
for the velocity field than in the body of the paper, namely allow the $\lambda$
factor to be proper-time dependent:
\bl{e:lambdatau}
\Omega=\lambda\z{\tau}\eta .
\ee
We denote the $\tau\pd{}{\tau}$ derivation operator by a comma, and if possible,
we suppress the $\tau$ argument of the functions in the notation. The $\lambda(\tau)=1$
case was investigated in Appendix A as well as in the body of the paper, so we 
assume $\lambda(\tau)\neq 1$ throughout this Appendix.

We will use a Taylor-expansion around $\eta=0$ for $\Omega$ and
the pressure (that is, for $Q=\rec{\kappa+1}\ln p$):
\bl{e:Q_series}
Q=A\z{\tau}+B\z{\tau}\eta+\frac{C\z{\tau}}{2}\eta^2+\frac{D\z{\tau}}{3}\eta^3+\frac{E\z{\tau}}{4}\eta^4+\dots ,
\ee
We have nothing more to do than to put this into \eqs{e:eul_Rin}{e:energy_Rin}, and
make the l.h.s. vanish order by order in powers of $\eta$. It can be done without any
essential difficulty, but it is a very long calculation. Here in this Appendix we only
consider the case when one makes an additional assumption for the pressure field, namely,
that it factorizes in terms of $\tau$ and $\eta$, as in the solutions presented in the
body of the paper:
\bl{e:p_factor}
p=F\z{\tau}H\z{\eta}
\ee
That is, for the Taylor expansion we have
\be
Q=A\z{\tau}+B\eta+\frac{C}{2}\eta^2+\frac{D}{3}\eta^3+\frac{E}{4}\eta^4+\dots ,
\ee
where $A\z{\tau}=\rec{\kappa+1}\ln F\z{\tau}$ and the other functions are constants
(denoted by the same letter), the expansion coefficients of $\rec{\kappa+1}\ln H\z{\eta}$. We will
need only the second order terms for the present calculation: we expand
\eq{e:eul_Rin} to $\c{O}\z{\eta^3}$, and \eq{e:energy_Rin} to $\c{O}\z{\eta^4}$.
The following auxiliary formulas are needed:
\[
\tanh\z{(\lambda-1)\eta}=\z{\lambda-1}\eta-\rec{3}\z{\lambda-1}^3\eta^3+\dots ,
\]
\[
\frac{\sinh\lambda\eta}{\cosh\z{(\lambda-1)\eta}}\rec{\sinh\eta}=\lambda-\rec{3}\lambda\z{\lambda-1}\z{\lambda-2}\eta^2+
\]
\[
+\rec{45}\lambda\z{\lambda-1}\z{\lambda-2}\z{6\lambda^2-12\lambda+7}\eta^4+\dots .
\]
\Eq{e:eul_Rin} thus gives
\bea
B&+&\kz{\lambda'+C+\z{\lambda-1}\z{A'+\lambda}}\eta+D\eta^2+ \nonumber \\
 &+&\z{E-\rec{3}\z{\lambda-1}^3\z{A'+\lambda}}\eta^3=0 ,
\eea
and \eq{e:energy_Rin} is (to second order in $\eta$) 
\bea
&\kz{\kappa A'+d\lambda}+\z{\lambda-1}\kappa B\eta+ \nonumber \\ 
+&\z{\lambda-1}\kz{\frac{d-1}{3}\lambda\z{2-\lambda}+\lambda'+\kappa C}\eta^2=0 . \label{e:energy_series}
\eea
We immediately see that $B=D=0$, and we obtain the following equations:
\bea
\kappa A'+d\lambda                                 &=&0 , \label{e:1} \\
\lambda'+C+\z{\lambda-1}\z{A'+\lambda}             &=&0 , \label{e:2} \\
\frac{d-1}{3}\lambda\z{2-\lambda}+\lambda'+\kappa C&=&0 , \label{e:3} \\
\rec{3}\z{\lambda-1}^3\z{A'+\lambda}-E             &=&0 . \label{e:4}
\eea
For brevity we did not write the $\c{O}\z{\eta^4}$ term in \eq{e:energy_series}, but we will use 
the equation, which follows from the condition that the $\c{O}\z{\eta^4}$ term vanish:
\bea
\frac{d-1}{45}\lambda\z{\lambda-1}\z{\lambda-2}\z{6\lambda^2-12\lambda+7}&+& \nonumber \\
+\z{\lambda-1}\kappa E-\z{\kappa C+\lambda'}\frac{\z{\lambda-1}^3}{3}&=&0 . \label{e:energy_O4}	
\eea
Now substituting $A'$ from \eq{e:1} into \eq{e:2}, and then $\lambda'$
from \eq{e:2} in \eq{e:3}, this last equation will contain only $\lambda$
(not its derivative), and the $C$ constant. So \eq{e:3} will yield an 
\emph{algebraic} equation for $\lambda$, that is, $\lambda$ must be
a constant, thus $\lambda'$ vanishes. It follows that $A'$ is also constant. So the constants are
\be
A'=-\frac{d}{\kappa}\lambda ,
\ee
\be
C=-\lambda\z{\lambda-1}\z{1-\frac{d}{\kappa}} ,
\ee
\be
E=\rec{3}\lambda\z{\lambda-1}^3\z{1-\frac{d}{\kappa}} ,
\ee
and from \eqs{e:3}{e:energy_O4} we have two equations containing only $\lambda$:
\bl{e:lkd}
\kappa-d=\frac{\z{d-1}\z{2-\lambda}}{3\z{\lambda-1}} ,
\ee
\bl{e:lalg}
2\z{\lambda-1}^3\z{\kappa-d}+\frac{d-1}{15}\z{\lambda-2}\z{6\lambda^2-12\lambda+7}=0 .
\ee
We excluded the trivial $\lambda=0$ case, and as mentioned before, the $\lambda=1$
Hwa-Bjorken case was treated separately. Substituting \eq{e:lkd} into \eq{e:lalg}
we obtain
\bl{e:lambda}
\z{d-1}\z{\lambda-2}\z{4\lambda^2-8\lambda+3}=0 .
\ee
It is now easy to see that there are only those solutions which we presented in the body of the
paper: if $d=1$, then \eq{e:lkd} implies that 
$\kappa=1$, but then \eq{e:lambda} is satisfied by  any value of the parameter $\lambda$.
Second, if $\lambda=2$, then \eq{e:lambda} is satisfied in any number of dimensions $d$, 
but \eq{e:lkd} implies a condition for the equation of state,  $\kappa=d$.
Third, the other two roots of \eq{e:lambda} are $\lambda=\frac{3}{2}$, and $\lambda=\rec{2}$,
and it is clear, that there are no other solutions of \eq{e:lkd}. 
So the only possibilities are those listed in Table~\ref{t:sol}. They are indeed
solutions, as can be verified straigthforwardly by substituting these
solutions to the hydrodynamical equations in Rindler coordinates. 
The conclusion of this Appendix is that we cannot
extend the investigated class of solutions in an easy way even if we allow
$\lambda$ to be $\tau$-dependent. 

%%%%%%%%%%%%%%%%%%%%%%%%%%%%%%%%%%%%%%%%%%%%%%%%%%%%%%%%%%%%%%%%%%%%%%%%%%%%%%%%%%%%%%%

\appendix
\section{Appendix C: Rapidity distributions}

In this Appendix we derive \eq{e:dndy-approx}, the formula which gives the
rapidity distribution of the produced particles for our new 1+1 dimensional
solutions, which are given by \eqs{e:vsol}{e:psol}, and the constants
are those in case $e.)$ in Table \ref{t:sol}. We assumed $\mu_B=0$,
that is $\varepsilon+p=T\sigma$. We utilized the following
expressions for $\sigma$ and $T$, similarly to \eqs{e:ssol}{e:Tsol}:
\bl{}
T=T_f\z{\frac{p}{p_f}}^{\rec{1+\kappa}} \sp
\sigma=\sigma_f\z{\frac{p}{p_f}}^{\frac{\kappa}{1+\kappa}} ,
\ee
so we assumed $\nu_\sigma(S)\equiv 1$ for the scaling function, and
$p(\tau,\eta)$ is given by \eq{e:psol}. Of course, $p_f=\sigma_fT_f$.
The subscript $_f$ means freeze-out. (The constants in \eqs{e:ssol}{e:Tsol} can always be
re-scaled, so we denoted the proper-time coordinate of
the freeze-out hypersurface at $\eta=0$ by $\tau_f$, and $p_f$, $T_f$ and $\sigma_f$
are the values of the thermodynamical quantities in this space-time point.)
Here, as in the body of the paper, $m$ is the mass of the produced particle,
and we parametrize the momentum of it as $k^{\mu}=\z{E,p}=m\z{\cosh y,\sinh y}$,
with $y$ being the rapidity.

We calculate the rapidity distribution in two cases. First, in the
fully $1+1$ dimensional case, when not only the flow but also the produced particles have
only $1$ degrees of freedom, as they propagate along the $z$ axis. Second, in the case, when
this $1+1$ dimensional solution is embedded in a $1+3$ dimensional space, and the produced
particles can have transverse momentum as well.

\subsection{The $1+1$ dimensional case}

The rapidity distribution is given by
\bl{}
\td{n}{y}=E\td{n}{p}=\int\m{d}^2x\,S(x^\mu,k^\mu) ,
\ee
where $S(x^\mu,k^\mu)$ is the source function, and the notation
$\m{d}^2x$ refers to the fact that we are in an $1+1$ dimensional space-time.
The explicit form of $S(x^\mu,k^\mu)$ is
\bl{}
S(x,k)=\frac{g}{2\pi\hbar}\exp\z{\frac{\mu(x)}{T(x)}-\frac{k_\mu u^\mu(x)}{T(x)}}k_\mu\m{d}\Sigma^\mu(x)\delta(x-x_f) ,
\ee
where the notations are: $u^\mu$ is the four-velocity of the flow, $\mu$ is the
chemical potential, $g$ is the spin-degeneracy factor, and $\m{d}\Sigma_\mu(x)$ is
the Cooper-Frye flux term. (The Cooper-Frye flux term, together with the delta
function, which selects the freeze-out hypersurface, is essentially the Lebesgue
vector-measure of the freeze-out hypersurface).

In our particular solution, $\mu(x)=0$, and $k_\mu u^\mu(x)=m\cosh\z{\lambda\eta-y}$.
We selected the freeze-out condition in a way to take advantage of the simplicity of this
relation: our condition is that the temperature at $\eta=0$ reaches a given
$T_f$ value, and the hyper-surface is pseudo-orthogonal to the four-velocity field, or 
in other words, $u^\mu(x)$ is parallel to $\m{d}\Sigma^\mu(x)$.
(This hypersurface is \emph{not} of constant temperature nor of constant $\tau$.
This choice is also motivated by the success of the Buda-Lund model in describing
particle spectra, see e.g.~\cite{Csanad:2004mm,Csanad:2005qr}.)
Another formulation of this condition is that the freeze-out hypersurface
is ,,locally synchronized'': an observer which moves together with the flow
sees this hypersurface as a constant time surface in its neighborhood.
The equation of this hypersurface is
\bl{e:hypersurface}
\z{\frac{\tau_f}{\tau}}^{\lambda-1}\cosh\z{(\lambda-1)\eta}=1 .
\ee
We can now calculate the Cooper-Frye term. We have for the integration
\[
\int\m{d}^2x\delta\z{x-x_f}k_\mu\m{d}\Sigma^\mu(x)\rightarrow
\]
\bl{}
\rightarrow\int_{-\infty}^\infty\m{d}\eta\,m\tau_f\cosh\z{\lambda\eta-y}\cosh^{\rec{\lambda-1}-1}\z{(\lambda-1)\eta} .
\ee
So the rapidity distribution is
\[
\td{n}{y}=\int_{-\infty}^\infty\m{d}\eta\frac{m\tau_f}{2\pi\hbar}\cosh\z{\lambda\eta-y}\cosh^{\rec{\lambda-1}-1}\z{(\lambda-1)\eta}\times
\]
\bl{}
\times\exp\z{-\frac{m}{T_f}\cosh\z{\lambda\eta-y}\cosh^\frac{\lambda}{\lambda-1}\z{(\lambda-1)\eta}} .
\ee
This integration can be calculated numerically, but an analytic approximation is also available.
The formula of the saddle-point integration is
\bl{e:saddle}
\int_{-\infty}^{\infty}\m{d}x\,f(x)g(x)\approx g(x_0)\sqrt{\frac{2\pi f^3(x_0)}{-f''(x_0)}} ,
\ee
for an $f(x)$ function, that has a sharp maximum peak at $x_0$, and a $g(x)$, which changes
smoothly around $x_0$. Essentially this is an approximation where one treats the smooth
function as a constant $g(x_0)$, and the sharp function as a Gaussian peak with parameters
obtained from $f(x_0)$ and $f''(x_0)$. This method can be applied very effectively when the
peaked function has a parameter, which rules its half-width, so this parameter can be
changed in such a way that $f(x)$ approximates a Dirac-delta. The more $f(x)$ is like
a Dirac-delta, the better the saddle-point approximation is (and in the limiting case it
becomes exact).

In our case, $f(\eta)$ is the exponential, $g(\eta)$
is the remaining, and $\frac{m}{T_f}$ is the parameter, which governs the half-width of $f(\eta)$.
For convenience, we introduce the $\alpha=\frac{2\lambda-1}{\lambda-1}$ notation.
We have
\bl{}
\eta_0=\frac{y}{2\lambda-1} \sp f(\eta_0)=\exp\kz{-\frac{m}{T_f}
\cosh^\rec{\alpha}\z{\frac{y}{\alpha}}} ,
\ee
\bl{} f''(\eta_0)=-f(\eta_0)\frac{m}{T_f}\lambda\z{2\lambda-1}\cosh^{\frac{\lambda}{\lambda-1}-1}
\z{\frac{y}{\alpha}} ,
\ee
\bl{}
g(\eta_0)=\frac{m\tau_f}{2\pi\hbar}\cosh^\rec{\lambda-1}
\z{\frac{y}{\alpha}} .
\ee
Putting all this together, we obtain (approximately)
\bl{e:rap1p1}
\td{n}{y}=\mathcal{N}\cosh^{\frac{\alpha}{2}-1}\z{\frac{y}{\alpha}}\exp\kz{-\frac{m}{T_f}\cosh^\alpha\z{\frac{y}{\alpha}}+\frac{m}{T_f}} ,
\ee
\bl{e:rap1p1_0}
\mathcal{N}=\sqrt{\frac{2\pi T_f}{m\lambda\z{2\lambda-1}}}\frac{m\tau_f}{2\pi\hbar}\exp\z{-\frac{m}{T_f}} .
\ee
This is the expression of $\td{n}{y}$ in the true $1+1$ dimensional case.
The normalization constant $\mathcal{N}$ is $\td{n}{y}\Big{|}_{y=0}$, the value of $\td{n}{y}$ at mid-rapidity.

\subsection{The $1+3$ dimensional, embedded case}

If we take the transverse dynamics into account, that is, we embed the
$1+1$ dimensional solution in a 3 dimensional space (with $v_x=v_y=0$ and
no pressure gradients in transverse directions), the previous
\r{e:rap1p1}--\r{e:rap1p1_0} formulas need to be modified. We
have (in 3 spatial dimensions) $\rec{\z{2\pi\hbar}^3}$
instead of $\rec{2\pi\hbar}$, and $m_T$ instead of $m$, where
$m_T=\sqrt{E^2-p_z^2}=\sqrt{m^2+p_x^2+p_y^2}$ is the transverse mass.
With these modifications, the above \r{e:rap1p1}--\r{e:rap1p1_0}
expressions have the following meaning:
\[
E\td{n}{^3\v{k}}=\frac{\m{d}n}{\m{d}y\,2\pi m_T\m{d}m_T}=\mathcal{N}\cosh^{\frac{\alpha}{2}-1}\z{\frac{y}{\alpha}}\times
\]
\bl{e:rap1p3m}
\times\exp\kz{-\frac{m_T}{T_f}\cosh^\alpha\z{\frac{y}{\alpha}}+\frac{m_T}{T_f}} ,
\ee
\bl{e:rap1p3m_0}
\mathcal{N}=\sqrt{\frac{2\pi T_f}{m_T\lambda\z{2\lambda-1}}}\frac{m_T\tau_f}{\z{2\pi\hbar}^3}\exp\z{-\frac{m_T}{T_f}} .
\ee
In order to obtain the rapidity distribution, we have to integrate over the transverse mass:
\[
\td{n}{y}=2\pi\int_m^\infty\m{d}m_T\,m_T \sqrt{\frac{2\pi T_f}{m_T\lambda\z{2\lambda-1}}}\frac{m_T\tau_f}{\z{2\pi\hbar}^3}\times
\]
\bl{}
\times\exp\kz{-\frac{m_T}{T_f}\cosh^\alpha\z{\frac{y}{\alpha}}}\cosh^{\frac{\alpha}{2}-1}\z{\frac{y}{\alpha}} .
\ee
Here $m$ is the real mass of the produced particles, e.\@g.\@ $m=140$ MeV for pions.
The integral over $m_T$ can be also calculated in the limiting case of $\frac{m}{T}\to\infty$.
(This is the same condition as that for the saddle-point integration to be exact.)
In this case, we can apply the simple approximation
\bl{}
\int_a^\infty h(x)\exp\z{-\beta x}\m{d}x \approx \frac{h(a)}{\beta}
\ee
for $a\gg\rec{\beta}$, and for any $h(x)$, that varies smoothly around $a$.
Our case is $h(m_T)=m_T^{3/2}$. So we have
\bl{e:rap1p3}
\td{n}{y}=\mathcal{N}'\cosh^{-\frac{\alpha}{2}-1}\z{\frac{y}{\alpha}}
\exp\kz{-\frac{m}{T_f}\cosh^\alpha\z{\frac{y}{\alpha}}} ,
\ee
\bl{e:rap1p3_0}
\mathcal{N}'=\sqrt{\frac{2\pi mT_f^3}{\lambda\z{2\lambda-1}}}\frac{A m\tau_f}{2\pi\hbar} ,
\ee
with $A$ being the transverse cross section of the fluid. These are the results quoted in \eq{e:dndy-approx}.

%%%%%%%%%%%%%%%%%%%%%%%%%%%%%%%%%%%%%%%%%%%%%%%%%%%%%%%%%%%%%%%%%%%%%%%%%%%%%%%%%%%%%%%

\appendix
\section{Appendix D: General solution for the $d=1$, $\kappa=1$ case, investigation of stability}

If $\varepsilon=\kappa p$, and we take the choice of $\kappa=1$, then the hydrodynamical
equations (that is, \eqs{e:3deul}{e:3denergy}) have an intrinsic simplicity. Let us write
them down in this case! In this Appendix we denote the derivation with respect to
$t$ by dotting, and the derivation with respect to $r$ by comma. So the Euler equation is
\bl{e:CR1}
\rec{1-v^2}\z{\dot{v}+vv'}=-\rec{2p}\z{p'+v\dot{p}} .
\ee
We have $\varepsilon+p=2p$ for the enthalpy density. The energy conservation is 
\be
\rec{2p}\z{\dot{p}+vp'}=-\rec{2\z{1-v^2}}\z{\z{v^2}\dot{}+v\z{v^2}{}'}-v' ,
\ee
or in another form 
\bl{e:CR2}
\rec{1-v^2}\z{v'+v\dot{v}}=-\rec{2p}\z{\dot{p}+vp'} .
\ee
We used the fact that in $d=1$ dimensions $\z{\nabla\v{v}}=v'$ simply.
Taking a proper linear combination of \eqs{e:CR1}{e:CR2} we have the following:
\bea
\rec{1-v^2}v'      &= \Omega'      &= -\rec{2p}\dot{p} , \\
\rec{1-v^2}\dot{v} &= \dot{\Omega} &= -\rec{2p}p' .
\eea
Here we again used the $\Omega=\m{arctanh}v$ notation. 
These equations are analogous to the Cauchy-Riemann equations known in complex analysis,
and the general solution of them is easily written down: it contains two arbitrary
functions, $F\z{t+r}$ and $G\z{t-r}$, one for each direction of wave-propagation:
\bea
\Omega\z{t,r}&=& \rec{2}\kz{G\z{t-r}+F\z{t+r}} , \\
\ln (p \z{t,r}/p_0 )&=&        \kz{G\z{t-r}-F\z{t+r}} .
\eea
Thus this case is a particular one: the \emph{general} solution of
the relativistic hydrodynamical equations is found, and then it can be easily
fitted to a given initial condition. If we prescribe $p\z{t(r),r}$ and $\Omega\z{t(r),r}$ on some
$t=t(r)$ Cauchy hypersurface, then the proper $F$ and $G$ functions can be found as detailed below, and then the
solution of the initial value problem is that the two wave-shapes propagate in the opposite direction.
As a first example, let us consider the $\Omega=\lambda\eta$ solution presented in the body of the paper.
It can be easily verified, that the following $F$ and $G$ functions indeed yield it:
\bea
F\z{t+r} &=  &\lambda\ln\z{\frac{t+r}{\tau_0p_0^{-1/2\lambda}}} , \\
G\z{t-r} &= -&\lambda\ln\z{\frac{t-r}{\tau_0p_0^{-1/2\lambda}}} .
\eea

To simplify the presentation of the general case, let us introduce
\bl{e:pilog}
\Pi(t,r) = \rec{2} \ln (p\z{t,r}/p_0),
\ee
and give the initial boundary conditions in the form of three functions,
the Cauchy hypersurface $t_0(r)$, the initial velocity field $v = \tanh(\Omega)$ 
given by the form of $\Omega(t_0(r), r)$ and the initial pressure distribution
is given by $ p = p_0 \exp( 2 \Pi(t_0(r),r) $.

Let us introduce the positive and the negative light-cone coordinates on
the initial Cauchy hypersurface as
\bl{e:xpm}
     x^\pm (r) = t_0(r) \pm r.
\ee
It is important to assume, that on the Cauchy hypersurface, these relations can be inverted,
and $r$ can be expressed as a function of both the positive and the negative light-cone
coordinates:
\bea
	r & =&  f(x^+), \\
	r & =&  g(x^-).
\eea
For example, if the initial Cauchy hypersurface is given by $\tau=\tau_0 = const$,
then the equation of this hypersurface in time and coordinate is
$t_0(r) = \sqrt{\tau_0^2 + r^2}$
and the relation between the coordinate $r$ and the light-cone coordinates,
\bl{e:xpm-spec}
     x^\pm (r)=t_0(r)\pm r =\sqrt{\tau_0^2+r^2}\pm r,
\ee
which relations can be inverted as
\bea
	r &=& f(x^+)\,=\, \frac{x^+}{2}-\frac{\tau_0^2}{2 x^+},\\
	r &=& g(x^-)\,=\,-\frac{x^-}{2}+\frac{\tau_0^2}{2 x^-}.
\eea
Then the general solution of the hydrodynamical equations in this case
simplifies in terms of $\Omega$ and $\Pi$ to the following
relations:
\bea
\Omega(t,r) &=&  \rec{2}\sz{\Omega(t_0(f(t+r)),f(t+r))+\Omega(t_0(g(t-r)),g(t-r))} + \\
	      & & +\rec{2}\sz{\Pi   (t_0(f(t+r)),f(t+r))-\Pi   (t_0(g(t-r)),g(t-r))},  \\
\Pi(t,r)    &=&  \rec{2}\sz{-\Omega(t_0(f(t+r)),f(t+r))+\Omega(t_0(g(t-r)),g(t-r))}- \\
		& & -\rec{2}\sz{\Pi   (t_0(f(t+r)),f(t+r))+\Pi(t_0(g(t-r)),g(t-r))} .
\eea	
A by-product of the treatment of this Appendix is, that any $1+1$ dimensional relativistic hydrodynamical
solution with $\varepsilon=p$ EoS (including our one presented in the body of the paper,
with $\Omega=\lambda\eta$) is always stable in terms of $\Omega$ and $\Pi$: if there is a perturbation in the initial
condition for $\Pi$ and $\Omega$, then this perturbation propagates in both directions, but
it does not grow exponentially. It is interesting to contrast this behaviour,
the stability of our solutions in terms of rapidity and logarithmic pressure,
to the numerically found instability of the famous Landau-Khalatnikov solution 
for the perturbation of their initial conditions. 
Such a stability of our new, general solutions makes them particularly
attractive for testing various algorithms that integrate the solutions of
relativistic hydrodynamics numerically.

\end{document}